\documentclass[onecolumn,draftclsnofoot,12pt]{IEEEtran}
\usepackage{amsfonts}
\usepackage{times}
\usepackage{graphicx}
\usepackage{latexsym}
\usepackage{dsfont}
\usepackage{amssymb}
\usepackage{amsmath}
\usepackage{cite}
\usepackage{verbatim}
\usepackage{subfigure}

\newtheorem{algorithm}{Algorithm}

\newcommand{\figref}[1]{{Fig.}~\ref{#1}}

\def\bb0{{\mathbb{0}}}

\def\ba{{\mathbf{a}}}
\def\bb{{\mathbf{b}}}

\def\bh{{\mathbf{h}}}

\def\bm{{\mathbf{m}}}

\def\br{{\mathbf{r}}}

\def\bv{{\mathbf{v}}}

\def\bx{{\mathbf{x}}}

\def\b0{{\mathbf{0}}}

\def\bA{{\mathbf{A}}}
\def\bB{{\mathbf{B}}}

\def\bG{{\mathbf{G}}}

\def\bI{{\mathbf{I}}}

\def\bR{{\mathbf{R}}}

\def\bbC{{\mathbb{C}}}

\def\bbE{{\mathbb{E}}}

\def\bbR{{\mathbb{R}}}

\def\cA{\mathcal{A}}

\def\cD{\mathcal{D}}

\def\cN{\mathcal{N}}

\def\sf0{{\mathsf{0}}}

\def\kron{\otimes}

\usepackage{epstopdf}
\usepackage{enumerate}
\usepackage{algorithmicx}
\usepackage{algorithm}
\usepackage{amsmath}
\usepackage[noend]{algpseudocode}
\usepackage{float}
\usepackage{hyperref}
\usepackage{color}
\usepackage{makeidx}
\usepackage{bbm}
\usepackage{graphicx}
\usepackage{cleveref}
\usepackage{url}
\usepackage{steinmetz}
\usepackage[algo2e,linesnumbered,lined,boxed,commentsnumbered]{algorithm2e}

\newcommand{\sref}[1]{{Section}~\ref{#1}}

\newcommand{\pinv}[1]{\ensuremath{#1^{\dagger}}} 	

\DeclareMathOperator*{\argmax}{arg\,max}

\DeclareMathOperator\Arg{\textit{arg}}

\def \bEpsi{\boldsymbol{\Psi}}

\def \rm {\mathrm}
\def \bpsi{\boldsymbol{\psi}}
\begin{document}

\title{Enabling Large Intelligent Surfaces with Compressive Sensing and Deep Learning}
\author{Abdelrahman Taha, Muhammad Alrabeiah, and Ahmed Alkhateeb\\  \thanks{The authors are with the School of Electrical, Computer and Energy Engineering, Arizona State University, (Emails: a.taha, malrabei, aalkhateeb@asu.edu).}}
\maketitle

\begin{abstract}
	
Employing large intelligent surfaces (LISs) is a promising solution for improving the coverage and rate of future wireless systems. These surfaces comprise a massive number of nearly-passive elements that interact with the incident signals, for example by reflecting them, in a smart way that improves the wireless system performance. Prior work  focused on the design of the LIS reflection matrices assuming full knowledge of the channels. Estimating these channels at the LIS, however, is a key challenging problem. With the massive number of LIS elements, channel estimation or reflection beam training will be associated with (i) huge training overhead if all the LIS elements are passive  (not connected to a baseband) or with (ii) prohibitive hardware complexity and power consumption if all the elements are connected to the baseband through a fully-digital or hybrid analog/digital architecture.  This paper proposes efficient solutions for these problems by leveraging tools from compressive sensing and deep learning. First, a novel LIS architecture based on \textit{sparse channel sensors} is proposed. In this architecture, all the LIS elements are passive except for a few elements that are active (connected to the baseband of the LIS controller). We then develop two solutions that design the LIS reflection matrices with negligible training overhead. In the first approach, we leverage compressive sensing tools to construct the channels at all the LIS elements from the channels seen only at the active elements. These full channels can then be used to design the LIS reflection matrices with no training overhead. In the second approach, we develop a deep learning based solution where the LIS learns how to optimally interact with the incident signal given the channels at the active elements, which represent the current state of the environment and transmitter/receiver locations. We show that the achievable rates of the proposed compressive sensing and deep learning solutions approach the upper bound, that assumes perfect channel knowledge, with negligible training overhead and with  less than $1\%$ of the elements being active. This highlights a promising solution for LIS systems from both energy efficiency and spectral efficiency perspectives. 

\end{abstract}

\clearpage
\section{Introduction} \label{sec:Intro}

Large Intelligent Surfaces (LISs) are envisioned as intrinsic components of beyond-5G wireless systems \cite{Puglielli2015,Hu2018, Faisal2019, Nadeem2019,Jung2018, DeCarvalho2019, Huang2018,Liaskos2018,Bjoernson2019,Sanguinetti2019}. In its core design concept,  an LIS realizes a continuous electromagnetically-active surface by stacking a massive number of radiating/sensing elements. These elements interact with the incident signals, for example by reflecting them, in a way that improves the coverage and rate of the wireless systems \cite{Puglielli2015,Hu2018}. This concept is further motivated by the possible energy-efficient implementation using nearly passive elements such as analog phase shifters \cite{Huang2018,Hum2014,Tan2018}. 
Prior work focused on developing efficient designs for the LIS reflection matrices and evaluating their coverage and rate gains \textit{assuming the existence of global channel knowledge}. 
\textbf{But how can these extremely large-dimensional channels be estimated if the LIS is implemented using only reflecting elements?} Obtaining this channel knowledge may require huge-and possibly prohibitive-training overhead which represents a main challenge for the operation of the LIS systems. To overcome this challenge, this paper proposes a novel LIS hardware architecture jointly with compressive sensing and deep learning based solutions that design the LIS reflection matrices with negligible training overhead.

\subsection{Prior Work}

LIS-assisted wireless communication is attracting increased interest in the last few years. In terms of the circuit implementations, LIS surfaces can be realized using nearly passive elements with reconfigurable parameters \cite{Huang2018}. Candidate designs include conventional reflect-arrays \cite{Hum2014,Tan2018}, and software-defined metamaterials \cite{Zhang2018,Liaskos2018} among others. Using these surfaces, several signal processing solutions have been proposed to optimize the design of the LIS reconfigurable parameters (mostly considering the LIS as a reflecting surface) \cite{Huang2018,Nadeem2019,Basar2019}. In \cite{Huang2018}, an LIS-assisted downlink multi-user setup is considered with single-antenna users. The LIS elements in \cite{Huang2018} are modeled as quantized phase shifters/reflectors and computational low-complexity algorithms were developed to design these LIS phase matrices. In \cite{Nadeem2019}, an LIS-assisted downlink scenario, similar to that in \cite{Huang2018}, was considered and the precoder matrix at the base station as well as the LIS reflection matrices were designed, focusing on the case where a line-of-sight (LOS) exists between the base station and the LIS. In \cite{Basar2019}, a new transmission strategy combining LIS with index modulation was proposed to improve the system spectral efficiency. In terms of the overall system performance,  \cite{Jung2018} considered an uplink multi-user scenario and characterized the data rates with channel estimation errors.  

\textbf{The Critical Challenge:}
All the prior work in \cite{Huang2018,Nadeem2019,Basar2019,Jung2018,Tan2018} assumed that the knowledge about the channels between the LIS and the transmitters/receivers is available at the base station, either perfectly or with some error. Obtaining this channel knowledge, however, is one of the most critical challenges for LIS systems due to the massive number of antennas  (LIS elements) and the hardware constraints on these elements. More specifically, if the LIS elements are implemented using phase shifters that just reflect the incident signals, then there are two main approaches for designing the LIS reflection matrix. 
The first approach is to estimate the LIS-assisted channels at the transmitter/receiver by training all the LIS elements, normally one by one, and then use the estimated channels to design the reflection matrix. This yields massive channel training overhead due to the very large number of elements at the LIS. Instead of the explicit channel estimation, the LIS reflection matrix can be selected from quantized codebooks via online beam/reflection training. This is similar to the common beam training techniques in mmWave systems that employ similar phase shifter architectures \cite{Wang2009,Hur2013}. To sufficiently quantize the space, however, the size of the reflection codebooks needs normally to be in the order of the number of antennas, which leads to huge training overhead. 
To avoid this training overhead, a trivial solution is to employ fully-digital or hybrid analog/digital architectures at the LIS, where every antenna element is connected somehow to the baseband where channel estimation strategies can be used to obtain the channels \cite{Alkhateeb2014d,HeathJr2016,Alkhateeb2014}. This solution, however, leads to high hardware complexity and power consumption given the massive number of LIS elements. 

\subsection{Contribution} 
In this paper, we consider an LIS-assisted wireless communication system and propose a novel LIS architecture as well as compressive sensing and deep learning based solutions  that design the LIS reflection matrix with negligible training overhead. More specifically, the contributions of this paper can be summarized as follows.
\begin{itemize}
	\item \textit{Novel LIS hardware architecture:} We introduce a new LIS architecture where all the elements are passive except a few randomly distributed active channel sensors. Only those few active sensors are connected to the baseband of the LIS controller and are used to enable the efficient design of the LIS reflection matrices with low training overhead.  
	
	\item \textit{Compressive sensing based LIS reflection matrix design:} Given the new LIS architecture with randomly distributed active elements, we develop a compressive sensing based solution to recover the full channels between the LIS and the transmitters/receivers from the \textit{sampled} channels sensed at the few active elements. Using the constructed channels, we then design the LIS reflection matrices with no training overhead. We show that the proposed solution can efficiently design the LIS reflection matrices when only a small fraction of the LIS elements  are active yielding a promising solution for LIS systems from both energy efficiency and training overhead perspectives. 
	
	\item \textit{Deep learning based LIS reflection matrix design:} Leveraging tools from machine/deep learning, we propose a solution that  learns the direct mapping from the sampled channels seen at the active LIS elements and the optimal LIS reflection matrices that maximize the system achievable rate. Essentially, the proposed approach teaches the LIS system how to interact with the incident signal given the knowledge of the sampled channel vectors, that we call \textit{environment descriptors}. In other words, the LIS learns that when it observes these environment descriptors, it should reflect the incident signal using this reflection matrix. Different than the compressive sensing solution, the deep learning approach leverages the prior observations at the LIS and does not require any knowledge of the array structure. 
\end{itemize}
The proposed solutions are extensively evaluated using the accurate ray-tracing based DeepMIMO dataset \cite{DeepMIMO2019}. The results show that the developed compressive sensing and deep learning solutions can approach the optimal upper bound, that assumes perfect channel knowledge, when only a few LIS elements are active and with almost no training overhead, highlighting potential solutions for LIS systems.

\textbf{Notation}: We use the following notation throughout this paper: $\bA$ is a matrix, $\ba$ is a vector, $a$ is a scalar, $\cA$ is a set of scalars, and $\boldsymbol{\mathcal{A}}$ is a set of vectors. $\|\ba \|_p$ is the p-norm of $\ba$. $|\bA|$ is the determinant of $\bA$, whereas $\bA^T$, $\bA^H$, $\bA^*$, $\bA^{-1}$, $\pinv{\bA}$ are its transpose, Hermitian (conjugate transpose), conjugate, inverse, and pseudo-inverse respectively. $[\bA]_{r,:}$ and $[\bA]_{:,c}$ are the $r$th row and $c$th column of the matrix $\bA$, respectively. $\mathrm{diag}(\ba)$ is a diagonal matrix with the entries of $\ba$ on its diagonal. $\bI$ is the identity matrix. $\mathbf{1}_{N}$ and $\mathbf{0}_{N}$ are the $N$-dimensional all-ones and all-zeros vector respectively. $\bA \odot \bB$ and $\bA \kron \bB$ are the Hadamard and Kronecker products of $\bA$ and $\bB$, respectively. $\cN(\bm,\bR)$ is a complex Gaussian random vector with mean $\bm$ and covariance $\bR$. $\bbE\left[\cdot\right]$ is used to denote expectation. $\Arg \left( \ba \right) $ is a vector of arguments of the complex vector $\ba$. $\boldsymbol{\mathrm{vec}}(\mathbf{A})$ is a vector whose elements are the stacked columns of matrix $\mathbf{A}$.

\section{System and Channel Models} \label{sec:SysCh Model}

In this section, we describe the adopted system and channel models for the large intelligent surfaces (LISs).

\subsection{System Model} \label{sec:Sys_Model}

\begin{figure}[t] \centerline{\includegraphics[scale=1]{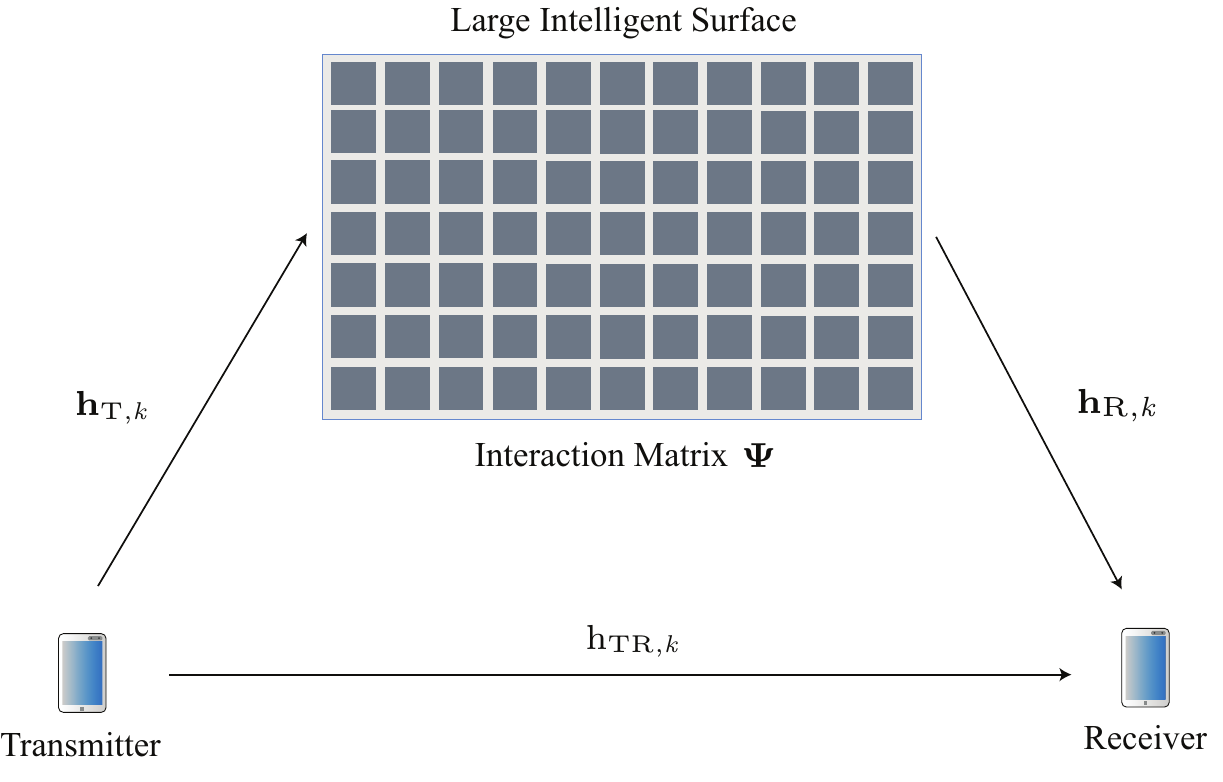}}
	\caption{This figure illustrates the system model where the transmitter-receiver communication is assisted by a large intelligent surface (LIS). The LIS is interacting with the incident signal through an interaction matrix $\boldsymbol{\Psi}$.} 
	\label{fig:Sys_Model}
\end{figure}


Consider the system model shown in \figref{fig:Sys_Model} where a transmitter is communicating with a receiver, and this communication is assisted by a large intelligent surface (LIS). For simplicity, we assume that the LIS has $M$ antennas while both the transmitter and receiver are single-antenna. The proposed solutions and results in this paper, however, can be extended to multi-antenna transceivers. Note also that these transmitters/receivers can represent either base stations or user equipment. 

Adopting an OFDM-based system of $K$ subcarriers, and defining $\bh_{\rm{T},k}, \bh_{\rm{R},k} \in \mathbb{C}^{M \times 1}$  as the  $M \times 1$ uplink channels from the transmitter/receiver to the LIS at the $k$th subcarrier,  $\bh_{\rm{T},k}^T, \bh_{\rm{R},k}^T$  as the downlink channels by reciprocity, $h_{\rm{TR},k} \in \mathbb{C}$ as the direct channel between the transmitter and receiver, then we can write the received signal at the receiver as 
\begin{equation}
y_k=\underbrace{\bh_{\rm{R},k}^T \bEpsi_k \bh_{\rm{T},k} s_k}_{\text{LIS-assisted link}} +  \underbrace{h_{\rm{TR},k} s_k}_\text{Direct link} + n_k, 
\end{equation}
where $s_k$ denotes the transmitted signal over the $k$th subcarrier, and satisfies $\bbE[\left|s_k\right|^2]=\frac{P_\rm{T}}{K}$, with $P_\rm{T}$ representing the total transmit power, and $n_k \sim \mathcal{N}_\mathbb{C}(0,\sigma_n^2)$ is the receive noise. The $M \times M$ matrix $\bEpsi_k$, that we call the LIS interaction matrix, represents the \textit{interaction} of the LIS with the incident (impinging) signal from the transmitter. The overall objective of the LIS is then to interact with the incident signal (via adjusting $\bEpsi_k$) in a way that optimizes a certain performance metric such as the system achievable rate or the network coverage. To simplify the design and analysis of the algorithms in this paper, we will focus on the case where the direct link does not exist. This represents the scenarios where the direct link is either blocked or has negligible receive power compared to that received through the LIS-assisted link. With this assumption, the receive signal can be expressed as
\begin{align}
y_k&=  \bh_{\rm{R},k}^T \bEpsi_k \bh_{\rm{T},k} s_k + n_k, \\
&\stackrel{(a)}{=}  \left( \bh_{\rm{R},k} \odot  \bh_{\rm{T},k}  \right)^T \boldsymbol{\psi}_k s_k + n_k,
\end{align}
where (a) follows from noting that the interaction matrix $\bEpsi_k$ has a diagonal structure, and denoting the diagonal vector as $\boldsymbol{\psi}_k$, i.e., $\bEpsi_k=\mathrm{diag}\left({\boldsymbol{\psi}_k}\right)$. This diagonal structure results from the operation of the LIS where every element $m, m= 1, 2, ..., M$ reflects the incident signal after multiplying it with an interaction factor $\left[\boldsymbol{\psi}_k\right]_m$. Now, we make two important notes on these interaction vectors. 
First, while the interaction factors, $\left[\boldsymbol{\psi}_k\right]_m, \forall m,k$, can generally have different magnitudes (amplifying/attenuation gains), it is more practical to assume that the LIS elements are implemented using only phase shifters. Second, since the implementation of the phase shifters is done in the analog domain (using RF circuits), the same phase shift will be applied to the signals on all subcarriers, i.e., $\bpsi_k=\bpsi, \forall k$. Accounting for these practical considerations, we assume that every interaction factor is just a phase shifter, i.e., $\left[\boldsymbol{\psi}\right]_m=e^{j \phi_m}$. Further, we will call the interaction vector $\bpsi$ in this case the \textit{reflection beamforming} vector. 

\subsection{Channel Model} \label{subsec:Ch_Model}
In this paper, we adopt a wideband geometric channel model for the channels $\bh_{\rm{T},k}, \bh_{\rm{R},k}$ between the transmitter/receiver and the LIS \cite{Alkhateeb2018}. Consider a transmitter-LIS uplink channel $\bh_{\rm{T},k}$ (and similarly for $\bh_{\rm{R},k}$) consisting of $L$ clusters, and each cluster $\ell$ is contributing with one ray of time delay $\tau_{\ell} \in \bbR$, a complex coefficient $\alpha_{\ell} \in \bbC$, and azimuth/elevation angles of arrival, $\theta_{\ell},\phi_{\ell} \in [0, 2 \pi)$.  Let $\rho_{\rm{T}}$ denote the path loss between the transmitter and the LIS and $p\left(\tau\right)$ characterizes the pulse shaping function for $T_{S}$-spaced signaling evaluated at $\tau$ seconds. The delay-$d$ channel vector, $\mathsf{\boldsymbol{h}}_{rd} \in \mathbb{C}^{M \times 1}$, between the transmitter and the LIS can then be defined as 
\begin{equation}
\mathsf{\boldsymbol{h}}_{\rm{T}, d} = \sqrt{\frac{M}{\rho_\rm{T}}} \sum_{ \ell =1}^L \alpha_{\ell} \ p(d T_S - \tau_{\ell}) \ \ba\left(\theta_{\ell}, \phi_{\ell} \right),
\label{eq:Ch_k}
\end{equation} 
where $\ba(\theta_{\ell},\phi_{\ell}) \in \mathbb{C}^{M \times 1}$ denotes the array response vector of the LIS at the angles of arrival $\theta_{\ell},\phi_{\ell}$. Given this delay-$d$ channel, the frequency domain channel vector at subcarrier $k$, $\bh_{\rm{T},k}$, can be written as 
\begin{equation} \label{eq:5}
\bh_{\rm{T}, k} = \sum_{d=0}^{D-1} \mathsf{\boldsymbol{h}}_{\rm{T}, d} \ e^{-j\frac{2\pi k}{K}d}.
\end{equation}

Considering a block-fading channel model, $\bh_{\rm{T},k}$ and $\bh_{\rm{R},k}$ are assumed to stay constant over the channel coherence time, denoted $T_{C}$, which depends on the user mobility and the dynamics of the environment among others. It is worth noting that the number of channel paths $L$ depends highly on the operational frequency band and the propagation environment. For example, mmWave channels normally consist of a small number of channel paths, $\sim$3-5 paths, \cite{Rappaport2013,Rappaport2014,Samimi2014}, while sub-6 GHz signal propagation generally experiences rich scattering resulting in channels with more multi-path components.

\section{Problem Formulation} \label{sec:Formulation}
The objective of this paper is to design the LIS interaction vector (reflection beamforming vector), $\bpsi$, to maximize the achievable rate at the receiver. Given the system and channel models in \sref{sec:SysCh Model}, this achievable rate can be written as 
\begin{align} \label{eq:9}
R &=\frac{1}{K}\sum_{k=1}^K  \log_2\left(1+\mathsf{SNR} \left| \bh_{\rm{R},k}^T \mathbf{\Psi} \bh_{\rm{T},k}   \right|^2 \right),\\
   &= \frac{1}{K}\sum_{k=1}^K  \log_2\left(1+\mathsf{SNR} \left| \left(\bh_{\rm{T},k} \odot \bh_{\rm{R},k}\right)^T \boldsymbol{\psi} \right|^2 \right) ,
\end{align}
where  $\mathsf{SNR}=\frac{P_{\rm{T}}}{K \sigma_n^2}$ denotes the signal-to-noise ratio. As mentioned in \sref{sec:Sys_Model}, every element in the LIS reflection beamforming vector, $\bpsi$, is implemented using an RF phase shifter. These phase shifters, however, normally have a quantized set of angles and can not shift the signal with any phase. To capture this constraint, we assume that the reflection beamforming vector $\bpsi$ can only be picked from a pre-defined codebook $\boldsymbol{\mathcal{P}}$. Every candidate reflection beamforming codeword in $\boldsymbol{\mathcal{P}}$ is assumed to be implemented using quantized phase shifters. With this assumption, our objective is then to find the optimal reflection beamforming vector $\bpsi^\star$ that solves 
\begin{align}
\bpsi^\star= \argmax_{\bpsi \in \boldsymbol{\mathcal{P}}}   \hspace{10pt} \sum_{k=1}^K  \log_2\left(1+\mathsf{SNR} \left| \left(\bh_{\rm{T},k} \odot \bh_{\rm{R},k}\right)^T \boldsymbol{\psi} \right|^2 \right), \label{eq:optV}
\end{align}
to result in the optimal rate $R^\star$ defined as
\begin{align}
R^\star= \max_{\bpsi \in \boldsymbol{\mathcal{P}}}   \hspace{10pt}\frac{1}{K} \sum_{k=1}^K  \log_2\left(1+\mathsf{SNR} \left| \left(\bh_{\rm{T},k} \odot \bh_{\rm{R},k}\right)^T \boldsymbol{\psi} \right|^2 \right). \label{eq:optR}
\end{align}

Due to the quantized codebook constraint and the time-domain implementation of the reflection beamforming vector, i.e., using one interaction vector $\bpsi$ for all subcarriers, there is no closed form solution for the optimization problem in \eqref{eq:optV}. Consequently, finding the optimal reflection beamforming vector for the LIS $\bpsi^\star$ requires an exhaustive search over the codebook $\boldsymbol{\mathcal{P}}$. 

\textbf{The main challenge:} As characterized in \eqref{eq:optV}, finding the optimal LIS interaction vector $\bpsi^\star$ and achieving the optimal rate $R^\star$ requires an exhaustive search over the codebook $\boldsymbol{\mathcal{P}}$. Note that the codebook size should normally be in the same order of the number of antennas to make use of these antennas. This means that a reasonable reflection beamforming codebook for LIS systems will probably have thousands of candidate codewords. With such huge codebooks, solving the exhaustive search in \eqref{eq:optV} is very challenging. More specifically, there are two main approaches for performing the search in \eqref{eq:optV}. 
\begin{itemize}
	\item \textbf{Full channel estimation with offline exhaustive search:} In this approach, we need to estimate the full channels between the LIS and the transmitter/receiver, $\bh_{\rm{T},k}, \bh_{\rm{R},k}$ and use it to find the optimal reflection beamforming vector by the offline calculation of  \eqref{eq:optV}. Estimating these channel vectors, however, requires the LIS to employ a complex hardware architecture that connects all the antenna elements to a baseband processing unit either through a fully-digital or hybrid analog/digital architectures \cite{HeathJr2016,Alkhateeb2014d}. Given the massive numbers of antennas at large intelligent surfaces, this approach can yield \textbf{prohibitive hardware complexity} in terms of the routing and power consumption among others. If the LIS is operated and controlled via a base station or an access point \cite{Huang2018}, then this channel estimation process can be done at these communication ends. This, however, assumes an orthogonal training over the LIS antennas, for example by activating one LIS antenna at a time, which leads to \textbf{prohibitive training overhead} given the number of antennas at the LIS.     
	
	\item \textbf{Online exhaustive beam training:} Instead of the explicit channel estimation, the best LIS beam reflection vector $\bpsi^\star$ can be found through an over-the-air beam training process. This process essentially solves the exhaustive search in \eqref{eq:optV} by testing the candidate interaction vectors $\bpsi \in \boldsymbol{\mathcal{P}}$ one by one. This exhaustive beam training process, however, incurs again \textbf{very large training overhead} at the LIS systems.   
\end{itemize}

Our objective in this paper is to enable large intelligent surfaces by addressing this main challenge. More specifically, our objective is to  enable LIS systems to approach the optimal achievable rate in \eqref{eq:optR} adopting \textbf{low-complexity hardware architectures} and requiring \textbf{low training overhead}. For this objective, we first propose a novel energy-efficient LIS transceiver architecture in \sref{sec:LIS_arch}. Then, we show in Sections \ref{sec:BL_Solution}-\ref{sec:DL_Solution} how to employ this LIS architecture to achieve near-optimal achievable rates with negligible training overhead via leveraging tools from compressive sensing and deep learning.

\section{Large Intelligent Surfaces with Sparse Sensors: A Novel Architecture} \label{sec:LIS_arch}

\begin{figure}[t] \centerline{\includegraphics[scale=1]{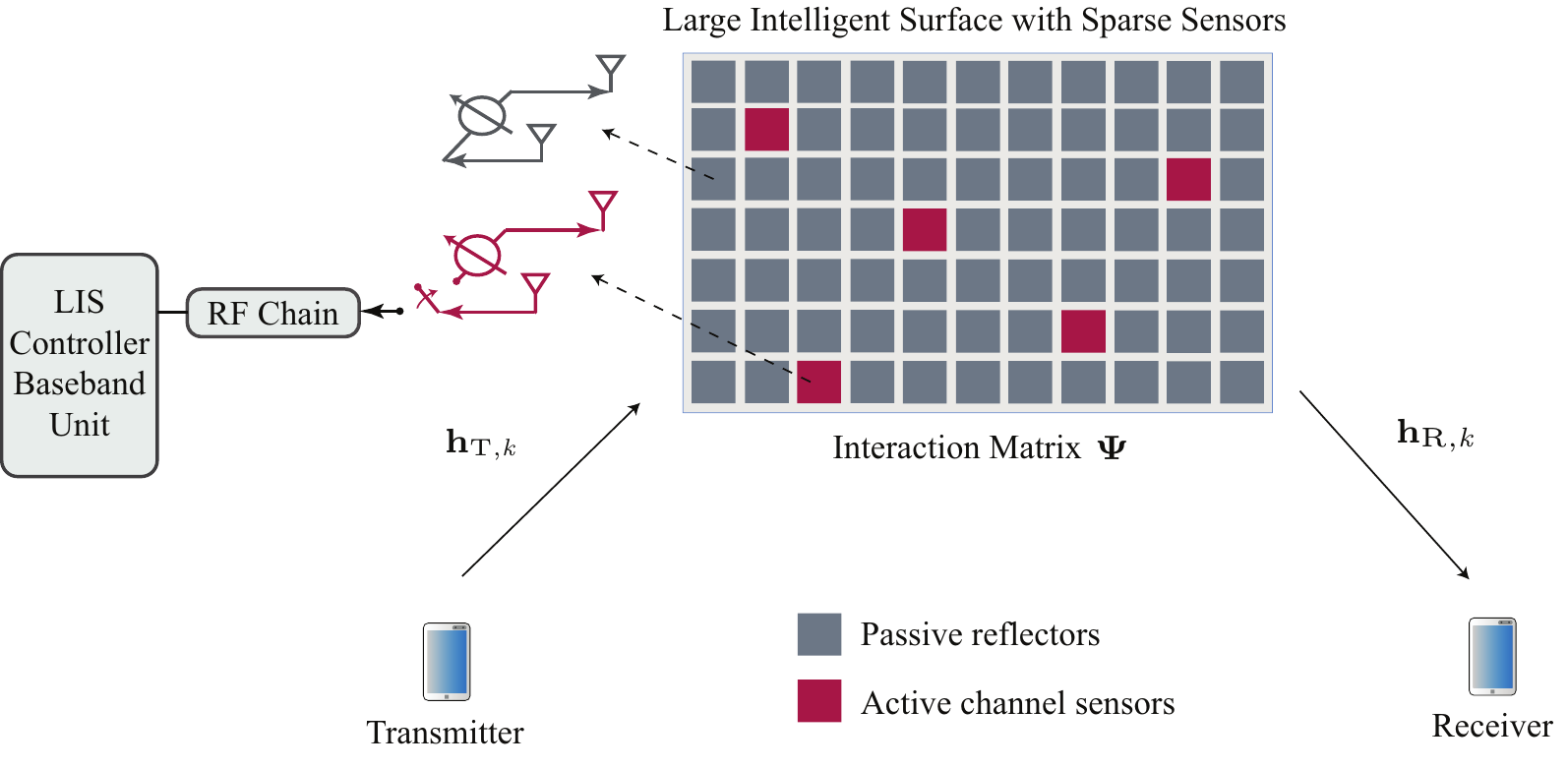}}
	\caption{This figure illustrates the proposed LIS architecture where $\overline{M}$ active channel sensors are randomly distributed over the LIS. These active elements have two modes of operation (i) a channel sensing mode where it is connected to the baseband and is used to estimate the channels and (ii) a reflection mode where it just reflects the incident signal by applying a phase shift. The rest of the LIS elements are passive reflectors and are not connected to the baseband.} 
	\label{fig:Sys_Model_Sparse}
\end{figure}

As discussed in \sref{sec:Formulation}, a main challenge for the LIS system operation lies in the  high hardware complexity and training overhead associated with designing the LIS interaction (reflection beamforming) vector, $\bpsi$. In order to overcome this challenge and enable LIS systems in practice, we propose the new LIS architecture in \figref{fig:Sys_Model_Sparse}. In this architecture, the LIS consists of (i) $M$ passive reflecting elements, each one is implemented using an RF phase shifter and is \textit{not connected} to the baseband unit, and (ii) a small number, $\overline{M} \ll M$, of \textit{active} channel sensors distributed over the large intelligent surface. For simplicity, we assume that the $\overline{M}$ active channel sensing elements are selected from the $M$ elements in the LIS and that they have two modes of operation (as shown in \figref{fig:Sys_Model_Sparse}): (i) A channel sensing mode where they work as receivers with full RF chains and baseband processing, and (ii) a reflection mode where they act just like the rest of the passive elements that reflect the incident signal. It is important to note  that while we describe the $M$ phase shifting elements as passive elements to differentiate them from the $\overline{M}$ active channel sensors, they are normally implemented using reconfigurable active RF circuits \cite{Hum2014,Foo2017}. Next, we define the channels from the transmitter/receiver to the active channel sensors of the LIS, and then discuss how to leverage this energy-efficient LIS architecture for designing the LIS interaction vector $\bpsi$.

\textbf{Sampled channel vectors:} We define the $\overline{M} \times 1$ uplink \textit{sampled} channel vector, $\overline{\bh}_{\rm{T},k} \in \mathbb{C}^{\overline{M} \times 1}$, as the channel vector from the transmitter to the $\overline{M}$ active elements at the LIS. This vector can then be expressed as 
\begin{equation} \label{eq:8}
\overline{\bh}_{\rm{T},k} = \bG_\rm{LIS} \ \bh_{\rm{T},k},
\end{equation}
where $\bG_\rm{LIS}$ is an $\overline{M} \times M$ selection matrix that selects the entries of the original channel vector, $\bh_{\rm{T},k}$, that correspond to the active LIS elements. If $\mathcal{A}$ defines the set of indices of the active LIS antenna elements, $\left|\mathcal{A}\right|=\overline{M}$, then $\bG_\rm{LIS}=[\bI]_{\mathcal{A},:}$, i.e., $\bG_\rm{LIS}$ includes the rows of the $M \times M$ identity matrix, $\bI$, that correspond to the indices of the active elements.  The sampled channel vector, $\overline{\bh}_{\rm{R},k} \in \mathbb{C}^{\overline{M} \times 1}$, from the receiver to the $\overline{M}$ active sensors of the LIS is similarly defined. 
Finally, we define $\overline{\mathbf{h}}_{k} = \overline{\mathbf{h}}_{\rm{T},k} \odot \overline{\mathbf{h}}_{\rm{R},k}$ as the overall LIS sampled channel vector at the $k$th subcarrier.

\textbf{Designing the LIS interaction vector:}  For the system model in \sref{sec:Sys_Model} with the proposed LIS architecture in \figref{fig:Sys_Model_Sparse}, estimating the sampled channel vectors $\overline{\bh}_{\rm{T},k}, \overline{\bh}_{\rm{R},k}$ can be easily done with a few pilot signals. For example, adopting an uplink training approach, the transmitter can send one pilot signal that will be simultaneously received and processed by all the active elements in the LIS to estimate $\overline{\bh}_{\rm{T},k}$ (and similarly for $\overline{\bh}_{\rm{R},k}$). Given these sampled channel vectors, however, how can the LIS find the optimal reflection beamforming vector $\bpsi^\star$ that solves \eqref{eq:optR}? In the next two sections, we propose two approaches for addressing this problem leveraging tools from compressive sensing (in \sref{sec:BL_Solution}) and deep learning (in \sref{sec:DL_Solution}).

\section{Compressive Sensing Based LIS Interaction Design}  \label{sec:BL_Solution}

As shown in \sref{sec:Formulation}, finding the optimal LIS interaction (reflection beamforming) vector $\bpsi^\star$ that maximizes the achievable rate with no beam training overhead requires the availability of the full channel vectors $\bh_{\rm{T},k}, \bh_{\rm{R},k}$. Estimating these channel vectors at the LIS, however, normally requires that every LIS antenna gets connected to the baseband processing unit through a fully-digital or hybrid architecture \cite{Mendez-Rial2016,Alkhateeb2014d,Alkhateeb2014}. This can massively increase the hardware complexity with the large number of antennas at the LIS systems. In this section, and adopting the low-complexity LIS architecture proposed in \sref{sec:LIS_arch}, we show that it is possible to recover the full channel vectors $\bh_{\rm{T},k}, \bh_{\rm{R},k}$ from the sampled channel vectors $\overline{\bh}_{\rm{T}, k} , \overline{\bh}_{\rm{R}, k}$ when the channels experience sparse scattering. This is typically the case in mmWave and LOS-dominant sub-6 GHz systems.

\subsection{Recovering Full Channels from Sampled Channels:} \label{subsec:CS_sol}
With the proposed LIS architecture in \figref{fig:Sys_Model_Sparse}, the LIS can easily estimate the \textit{sampled} channel vectors $\overline{\bh}_{\rm{T}, k} , \overline{\bh}_{\rm{R}, k}$ through uplink training from the transmitter and receiver to the LIS with a few pilots. Next, we explain how to use these sampled channel vectors to estimate the full channel vectors $\bh_{\rm{T}, k}, \bh_{\rm{R}, k}$. First, note that the $\bh_{\rm{T}, k}$ in \eqref{eq:Ch_k}, \eqref{eq:5} (and similarly for $\bh_{\rm{R}, k}$) can be written  as
\begin{align}
\bh_{\rm{T}, k} &= \sqrt{\frac{M}{\rho_\rm{T}}} \sum_{d=0}^{D-1} \sum_{ \ell =1}^L \alpha_{\ell}  \ p(d T_S - \tau_{\ell}) \ \ba\left(\theta_{\ell}, \phi_{\ell} \right) \ e^{-j\frac{2\pi k}{K}d}, \\
&= \sum_{ \ell =1}^L  \beta_{\ell,k} \ \ba\left(\theta_{\ell}, \phi_{\ell} \right), 
\end{align}
where $\beta_{\ell,k}= \sqrt{\frac{M}{\rho_\rm{T}}} \alpha_\ell  \sum_{d=0}^{D-1}  p(d T_S - \tau_{\ell})  e^{-j\frac{2\pi k}{K}d}$.  Further, by defining the array response matrix $\bA$ and the $k$th subcarrier path gain vector $\boldsymbol{\beta}$ as
\begin{align}
\bA &=\left[ \ba\left(\theta_{1}, \phi_{1} \right),  \ba\left(\theta_{2}, \phi_{2} \right) ...,  \ba\left(\theta_{L}, \phi_{L} \right)\right], \\
\boldsymbol{\beta}_k &= \left[ \beta_{1,k}, \beta_{2,k}, ..., \beta_{L,k}\right]^T,
\end{align}
we can write $\bh_{\rm{T}, k}$  in a more compact way as $\bh_{\rm{T}, k} = \bA \ \boldsymbol{\beta}$. Now, we note that in several important scenarios, such as mmWave and LOS-dominant sub-6 GHz, the channel experiences sparse scattering, which results is a small number of paths $L$ \cite{Rappaport2014,HeathJr2016}. In order to leverage this sparsity, we follow \cite{Alkhateeb2014} and define the dictionary of array response vectors $\bA_\text{D}$, where every column constructs an an array response vector in one quantized azimuth and elevation direction. For example, if the LIS adopts a uniform planar array (UPA) structure, then we can define $\bA_\text{D}$ as
\begin{equation}
\bA_\text{D}=\bA^\rm{Az}_{\rm{D}} \kron \bA^\rm{El}_{\rm{D}}
\end{equation}
with $\bA^\rm{Az}_{\rm{D}}$ and $\bA^\rm{El}_{\rm{D}}$ being the dictionaries of the azimuth and elevation array response vectors. Every column in $\bA^\rm{Az}_{\rm{D}}$ (and similarly for $\bA^\rm{El}_{\rm{D}}$ ) constructs an azimuth array response in one quantized azimuth (elevation) direction. If the number of grid points in the azimuth and elevation dictionaries are $N_\rm{D}^\rm{Az}$ and $N_\rm{D}^\rm{El}$, respectively, and the number of horizontal and vertical elements of the UPA are $M_\mathrm{H}, M_\rm{V}$, where $M=M_\mathrm{H} M_\rm{V}$, then $\bA_\text{D}$ has dimensions $M \times N_\rm{D}^\rm{Az} N_\rm{D}^\rm{El}$. Now, assuming that size of the grid is large enough such that the  azimuth and elevation angles $\theta_\ell, \phi_\ell, \forall \ell$ matches exactly $L$ points in this grid (which is a common assumption in the formulations of the sparse channels estimation approaches \cite{Alkhateeb2014,HeathJr2016,Lee2014}), then we can rewrite $\bh_{\rm{T}, k}$ as
\begin{equation} \label{eq:construct_h}
\bh_{\rm{T}, k} = \bA_\rm{D} \ \bx_{\boldsymbol{\beta}},
\end{equation}
where $\bx_{\boldsymbol{\beta}}$ is an $N_\rm{D}^\rm{Az} N_\rm{D}^\rm{El}$ sparse vector with $L \ll N_\rm{D}^\rm{Az} N_\rm{D}^\rm{El}$ non-zero entries equal to the elements of $\boldsymbol{\beta}$. Further, these non-zero entries are in the positions that correspond to the channel azimuth/elevation angles of arrival. Next, let $\widehat{\overline{\bh}}_{\rm{T}, k}$ denote the noisy sampled channel vectors, then we can write
\begin{align}
\widehat{\overline{\bh}}_{\rm{T}, k} & = \bG_{\rm{LIS}} \bh_{\rm{T},k} + \bv_k, \\
& =  \bG_{\rm{LIS}}\bA_\rm{D} \ \bx_{\boldsymbol{\beta}} + \bv_k, \\
& = \boldsymbol{\Phi} \ \bx_{\boldsymbol{\beta}} + \bv_k,
\end{align}
where $\bv_k \sim \mathcal{N}_\mathbb{C} \left(\boldsymbol{0}, \sigma_n^2 \bI \right)$  represent the receive noise vector at the LIS active channel sensors and $\bG_\rm{LIS}$ is the selection matrix defined in \eqref{eq:8}. Now, given the equivalent sensing matrix, $\boldsymbol{\Phi}$ and the noisy sampled channel vector $\widehat{\overline{\bh}}_{\rm{T}, k}$, the objective is to estimate the sparse vector $\bx_{\boldsymbol{\beta}}$ that solves the non-convex combinatorial problem 
\begin{equation} \label{eq:sparse_h}
\min \left\|\bx_{\boldsymbol{\beta}}\right\|_0    \  \  \  \text{s.t.}  \ \left\| \widehat{\overline{\bh}}_{\rm{T}, k}  - \boldsymbol{\Phi} \ \bx_{\boldsymbol{\beta}}\right\|_2 \leq \sigma.
\end{equation}

Given the sparse formulation in \eqref{eq:sparse_h}, several compressive sensing reconstruction algorithms, such as orthogonal matching pursuit (OMP) \cite{Cai2011a,Tropp2004a}, can be employed to obtain an approximate solution for $\bx_{\boldsymbol{\beta}}$. With this solution for $\bx_{\boldsymbol{\beta}}$, the full  channel vector $\bh_{\rm{T},k}$ can be constructed according to \eqref{eq:construct_h}. Finally, the constructed full channel vector can be used to find the LIS reflection beamforming vector $\bpsi$ via an offline search using \eqref{eq:optV}. 

In this paper, we assume for simplicity that the $\overline{M}$ active channel sensors are randomly selected from the $M$ LIS elements, assuming that all the elements are equally likely to be selected. It is important, however, to note that the specific selection of the active elements designs the compressive sensing matrix $\boldsymbol{\Phi}$ and decides its properties. Therefore, it is interesting to explore the optimization of the active element selection, leveraging tools from nested arrays \cite{Pal2010}, co-prime arrays \cite{Vaidyanathan2011,Tan2014}, incoherence frames \cite{Rusu2018}, and difference sets \cite{Giannakis2005,Mendez-Rial2016}.

\subsection{Simulation Results and Discussion:} \label{subsec:CS_results}
\begin{figure}[t] \centerline{\includegraphics[scale=.6]{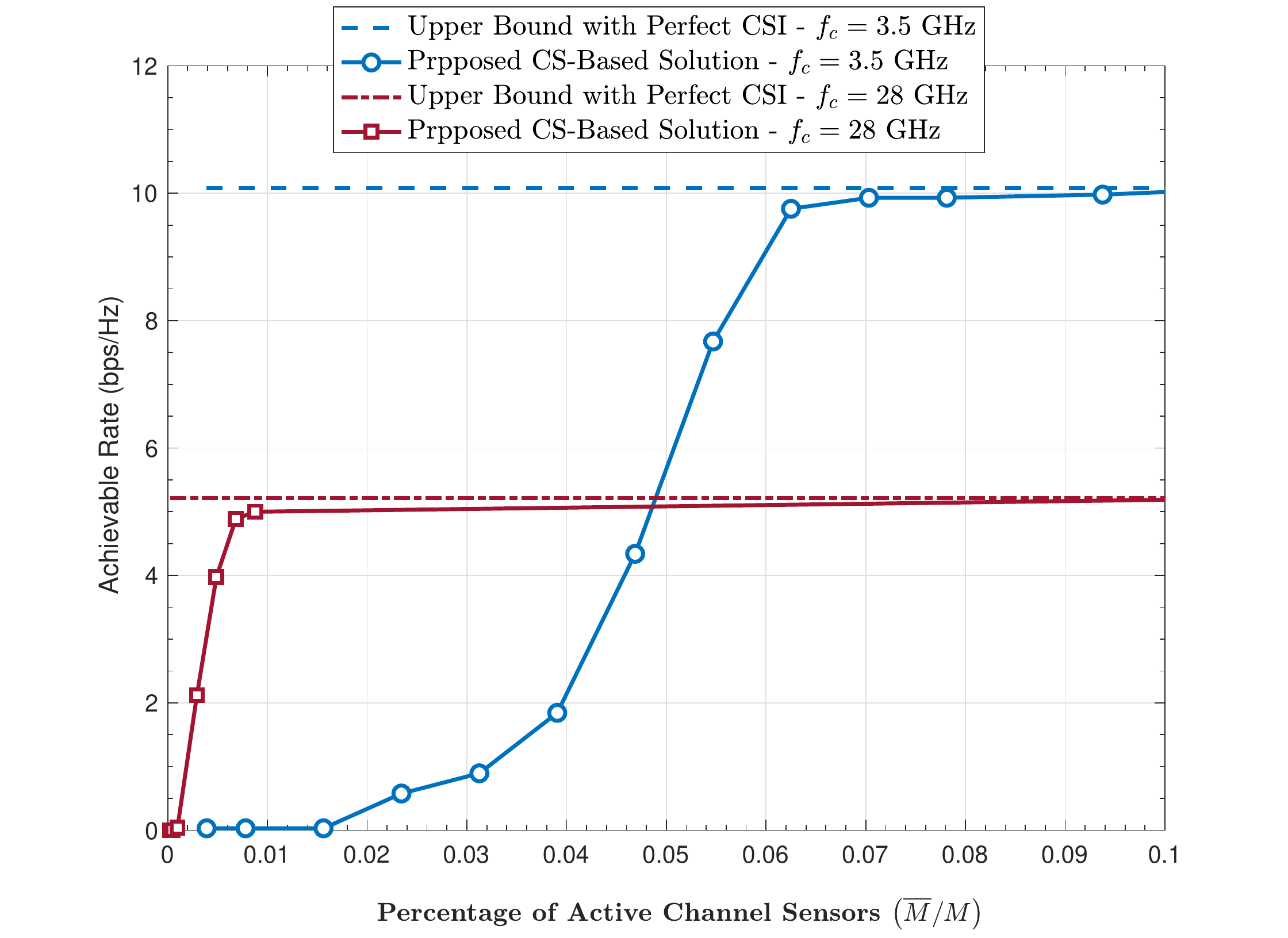}}
	\caption{ This figure plots the achievable rates using the proposed compressive sensing based solution for two scenarios, namely a mmWave 28GHz scenario and a low-frequency 3.5GHz one. These achievable rates are compared to the optimal rate $R^\star$ in \eqref{eq:optR} that assumes perfect channel knowledge. The figure illustrates the potential of the proposed solutions that approach the upper bound while requiring only a small fraction of the total LIS elements to be active.} 
	\label{fig:CS}
\end{figure}

To evaluate the performance of the proposed compressive sensing based solution, we consider a simulation setup at two different carrier frequencies, namely 3.5GHz and 28GHz. The simulation setup consists of one large intelligent surface with a uniform planar array (UPA) in the y-z plane, which reflects the signal coming from one transmitter to another receiver, as depicted in \figref{fig:Setup}. This UPA consists of $16 \times 16$ antennas at 3.5GHz and $64 \times 64$ antennas at 28GHz. We generate the channels using the publicly-available ray-tracing based DeepMIMO dataset \cite{DeepMIMO2019}, with the 'O1' scenario that consists of a street and buildings on the sides of the street. Please refer to \sref{subsec:Sim_Setup} for a detailed description of the simulation setup and  its parameters. 

Given this described setup, and adopting novel LIS architecture in \figref{fig:Sys_Model_Sparse}, we apply the proposed compressive-sensing based solution described in \sref{subsec:CS_sol} as follows: (i) We obtain the channel vectors $\bh_{\rm{T},k}, \bh_{\rm{R},k}$ using the ray-tracing based DeepMIMO dataset, and add noise with the noise parameters described in \sref{subsec:Sim_Setup}. (ii) Adopting the LIS architecture in \figref{fig:Sys_Model_Sparse}, we randomly select $\overline{M}$ elements to be active and construct the sampled channel vectors $\widehat{\overline{\bh}}_{\rm{T}, k}, \widehat{\overline{\bh}}_{\rm{R}, k}$. (iii) Using OMP with a grid of size $N_\rm{D}^\rm{Az} N_\rm{D}^\rm{El}, N_\rm{D}^\rm{Az}=2 M_\rm{H}, N_\rm{D}^\rm{El}=2 M_\rm{V}$, we recover an approximate solution of the full channel vectors and use this to search for the optimal LIS interaction vector using \eqref{eq:optV}. The achievable rate using this proposed compressive sensing based solution is shown in \figref{fig:CS} compared to the upper bound with perfect full channel knowledge, calculated according to \eqref{eq:optR}.

\textbf{Gains and Limitations:} In \figref{fig:CS}, we plot the achievable rates of the proposed compressive sensing based solution and upper bound versus the ratio of the active elements to the total number of antennas, i.e., $\overline{M}/M$. As shown in this figure, the proposed novel LIS architecture with the compressive sensing based solution can achieve almost the optimal rate with a small fraction of the LIS antennas being active. This illustrates the significant saving in power consumption that can be achieved using the LIS architecture in \figref{fig:Sys_Model_Sparse} that includes a few active channel sensors. Further, since the LIS reflection beamforming vector $\bpsi$ is obtained through an offline search with no beam training, the proposed solution approaches the optimal rate with negligible training overhead, ideally two uplink pilots to estimate $\widehat{\overline{\bh}}_{\rm{T}, k}, \widehat{\overline{\bh}}_{\rm{R}, k}$. This enables the proposed LIS systems to support highly-mobile applications such as vehicular communications and wireless virtual/augmented reality. 

Despite this interesting gain of the proposed compressive sensing based solution, it has a number of limitations. First, recovering the full channel vectors from the sampled ones according to \sref{subsec:CS_sol} requires the knowledge of the array geometry and is hard to extend to LIS systems with unknown array structures. Second, the compressive sensing solution relies on the sparsity of the channels and its performance becomes limited in scenarios with rich NLOS scattering. This is shown in \figref{fig:CS} as the compressive sensing based solution requires a higher ratio of the LIS elements to be active to approach the upper bound in the 3.5GHz scenario that has more scattering than the mmWave 28GHz case. Further, the compressive sensing solution does not leverage the previous observations to improve the current channel recovery. These limitations motivate the deep learning based solution that we propose in the following section.

\section{Deep Learning Based LIS Interaction Design}  \label{sec:DL_Solution}

In this section, we introduce a \textit{novel} application of deep learning in the reflection beamforming design problem of large intelligent surfaces. The section is organized as follows: First, the key idea of the proposed deep learning (DL) based reflection beamforming design is explained. Then, the system operation and the adopted deep learning model  are diligently described. We refer the interested reader to \cite{Goodfellow2016}
 for a brief background on deep learning.

\subsection{The Key Idea} \label{subsec:keyIdea}

\begin{figure}[t] 
	\includegraphics[width = \linewidth]{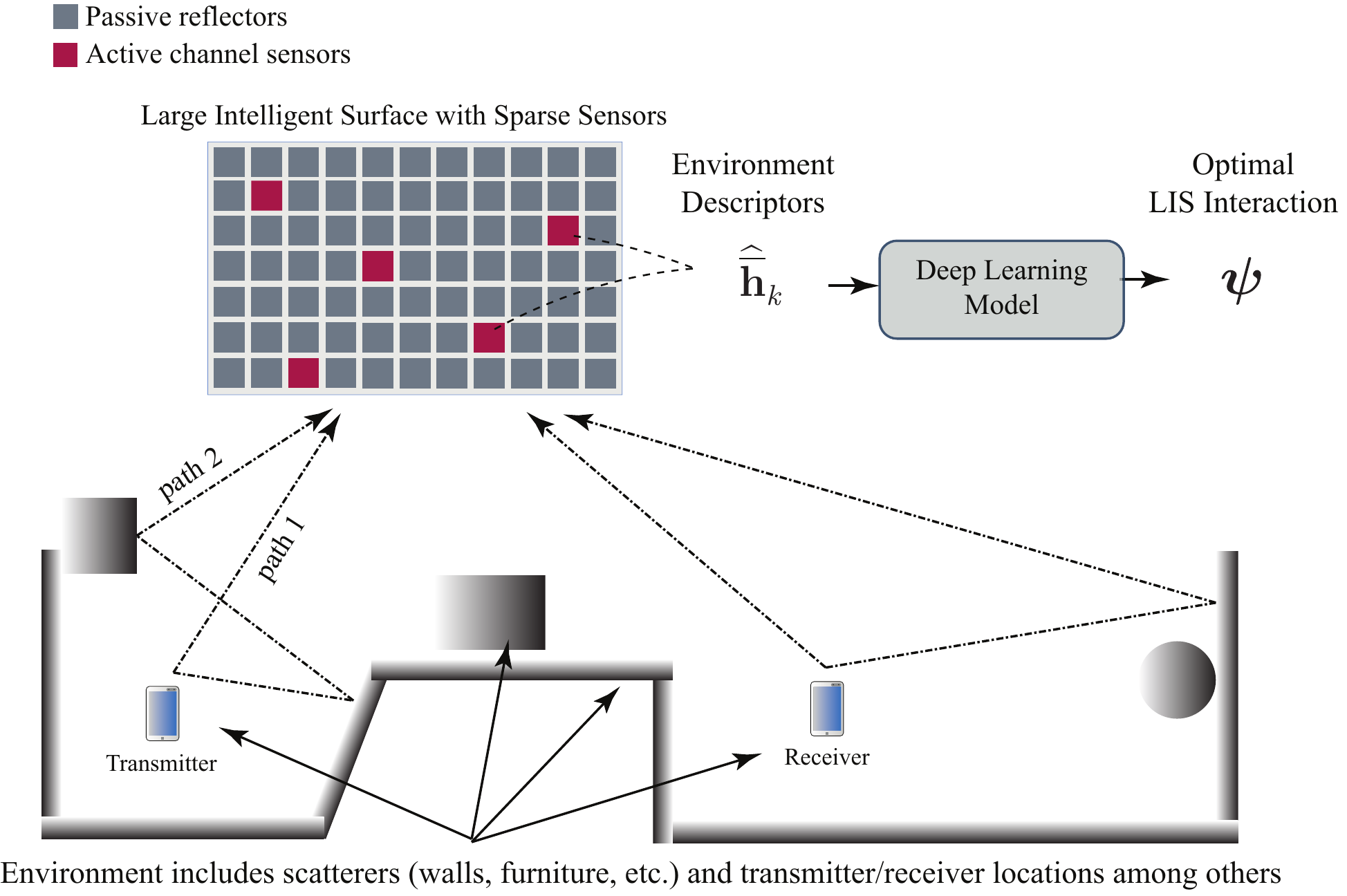}
	\caption{This figure summarizes the key idea of the proposed deep learning solution. The sampled channel vectors are considered as environment descriptors as they define, with some resolution, the transmitter/receiver locations and the surrounding environment. The deep learning model learns how to map the observed environment descriptors to the optimal LIS reflection vector. }
	\label{fig:Key_idea}
\end{figure}

The large intelligent surfaces are envisioned as key components of future networks \cite{Huang2018}. These surfaces will interact with the incident signals, for example by reflecting them, in a way that improves the wireless communication performance. In order to decide on this interaction, however, the LIS systems or their operating base stations and access points need to acquire some knowledge about the channels between the LIS and the transmitter/receiver. As we explained in \sref{sec:Formulation}, the massive number of antennas at these surfaces makes obtaining the required channel knowledge associated with (i) prohibitive training overhead if all the LIS elements are passive or (ii) infeasible hardware complexity/power consumption in the case of fully-digital or hybrid based LIS architectures.

The channel vectors/matrices, however, are intuitively some functions of the various elements of the surrounding environment such as the geometry, scatterer materials, and the transmitter/receiver locations among others. 
Unfortunately, the nature of this function---its dependency on the various components of the environment---makes its mathematical modeling very hard and infeasible in many cases. 
This dependence though means that the interesting role the LIS is playing could be enabled with some form of awareness about the surrounding environment. With this motivation, and adopting the proposed LIS architecture in \figref{fig:Sys_Model_Sparse}, we propose to consider the sampled channels seen by the few active elements of the LIS as \textit{environment descriptors} capturing a multi-path signature \cite{Alkhateeb2018,Alkhateeb2018b,Li2018b}, as shown in \figref{fig:Key_idea}. We then adopt deep learning models to learn the mapping function from these environment descriptors to the optimal LIS interaction (reflection beamforming) vectors. \textbf{In other words, we are teaching the LIS system how to interact with the wireless signal given the knowledge of the environment descriptors (sampled channel vectors).} It is worth emphasizing here that the sampled channel vectors can be obtained with negligible training overhead as explained in \sref{sec:LIS_arch}. In the ideal case, the learning model will be able to predict the optimal interaction vector given the environment descriptors. Achieving this means that the LIS system can approach the optimal rate in \eqref{eq:optR} with negligible training overhead (that is required to obtain the sampled channel vectors) and with low-complexity architectures (as only a few elements of the LIS are active).

\subsection{Proposed System Operation} \label{subsec:Sys_Oper}

In this section, we describe the system operation of the proposed deep learning based LIS interaction approach. The proposed system operates in two phases, namely (I) the learning phase and (II) the prediction phase. \\

\LinesNotNumbered 
\begin{algorithm}[t]
	\textbf{PHASE I: Learning phase} \\
	\For{ $s=1$ \KwTo  $S$} 
	{
		\tcp{For every channel coherence block $s$.}
		\nl  \label{Alg_H_bar_estI}LIS receives two pilots from the transmitter and receiver and estimates  $\widehat{\overline{\bh}}(s)$.
		
		\For{ $n=1$ \KwTo  $\left|\boldsymbol{\mathcal{P}}\right|$ }{
			\nl \label{Alg_ES1}LIS Reflects using $\boldsymbol{\psi}_n$ beam and receives the feedback $R_{n}(s)$.
			
		}

		\nl \label{DSx}LIS constructs $\br(s)=[R_{1}(s), R_2(s), ..., R_{|\boldsymbol{\mathcal{P}}|}(s)]$.
		
		\nl \label{DS}A new data point is added to the learning dataset: $\cD \leftarrow \left\langle \widehat{\overline{\bh}}(s), \br_{s} \right\rangle$.
		
		\nl \label{DT_2}LIS uses the interaction vector $\boldsymbol{\psi}_{n^\star}$, with $n^\star=\argmax_n R_n(s)$, for the data 
		reflection.
		
	}
	
	\nl \label{DLTrain1} Train the DL model using the available dataset $\cD$.

	\textbf{PHASE II: Prediction phase} \\
	\While{True}{
		
		\tcp{Repeat every channel coherence block.}
		
		\nl \label{Alg_H_bar_estII}LIS receives two pilots from the transmitter and receiver, and estimates $\widehat{\mathbf{\overline{h}}}$.
		
		\nl \label{Pred}LIS predicts the interaction (reflection) vector using the trained DL model.
		
		\nl \label{DT_2_II}LIS uses the interaction vector $\boldsymbol{\psi}_{n^\rm{DL}}$, with $n^\rm{DL}=\argmax_n \hat{R}_n$, for the data 
		reflection.
	}
	\caption{Proposed Deep Learning Based Reflection Beamforming}
	\label{alg1}
\end{algorithm}

\textbf{PHASE I: Learning phase} 
In this phase, the LIS employs an exhaustive search reflection beamforming approach as will be explained shortly while it is collecting the dataset for the deep learning model. Once the dataset is fully acquired, the LIS trains the deep learning model. 
Let the term ``data sample" indicate the data captured in one coherence block, and define the concatenated sampled channel vector $\overline{\mathbf{h}} = \textbf{vec} \left(  \left[ \overline{\mathbf{h}}_{1}, \overline{\mathbf{h}}_{2}, \dotsc, \overline{\mathbf{h}}_{K} \right] \right)$. Further, let $\overline{\mathbf{h}}(s)$ denote the concatenated sampled channel vector at coherence block $s, s=1, ..., S$. 
As depicted in Algorithm \ref{alg1}, at every coherence block $s$, the proposed LIS system operation consists of four steps, namely estimating the sampled channel vector, exhaustive beam training, constructing a new data sample for the learning dataset, and data transmission. After collecting the dataset, the deep learning model is trained. Next, we describe these steps in detail.
\begin{enumerate}
	\item \textit{Sampled channel estimation (line \ref{Alg_H_bar_estI}):} For every channel coherence block $s$, the transmitter and receiver transmits two orthogonal uplink pilots. The active elements of the LIS will receive these pilots and estimate the sampled channel vectors 
	\begin{equation}
	\widehat{\overline{\bh}}_{\rm{T}, k}(s) =  \overline{\bh}_{\rm{T},k}(s) + \bv_k, 
	\label{eq:noisy_ch}
	\end{equation}
	where $\bv_k \sim \mathcal{N}_\mathbb{C} \left(\boldsymbol{0}, \sigma_n^2 \bI \right)$ represent the receive noise vector at the LIS active channel sensors. The receiver-LIS sampled channel vector $\widehat{\overline{\bh}}_{\rm{R}, k}(s)$ will be similarly estimated. Finally, the vectors $\widehat{\overline{\mathbf{h}}}_{k}(s) = \widehat{\overline{\mathbf{h}}}_{\rm{T},k}(s) \odot \widehat{\overline{\mathbf{h}}}_{\rm{R},k}(s)$ and $\widehat{\overline{\mathbf{h}}}(s) = \textbf{vec} \left(  \left[ \widehat{\overline{\mathbf{h}}}_{1}(s), \widehat{\overline{\mathbf{h}}}_{2}(s), \dotsc, \widehat{\overline{\mathbf{h}}}_{K}(s) \right] \right)$ will be constructed.

	\item \textit{Exhaustive search over reflection beamforming codewords (line \ref{Alg_ES1}):} 
	In this step, the LIS performs an exhaustive beam training using the interaction/reflection codebook $\boldsymbol{\mathcal{P}}$. Particularly, the LIS attempts every candidate reflection beamforming vector, $\boldsymbol{\psi}_{n}, n=1, ..., |\boldsymbol{\mathcal{P}}|$, and receives a feedback from the receiver indicating the achievable rate using this interaction vector, $R_{n}{(s)}$, defined as 
	\begin{equation} 
	R_{n}^{(s)} = \frac{1}{K}\sum_{k=1}^K  \log_2\left(1+\mathsf{SNR} \left| \left(\bh_{\rm{T},k}(s) \odot \bh_{\rm{R},k}(s)\right)^T \boldsymbol{\psi}_n \right|^2 \right).
	\end{equation} \label{eq:23}
	Note that, in practice, the computation and feedback of the achievable rate $R_{n}^{(s)}$ will have some error compared to \eqref{eq:23} due to the limitations in the pilot sequence lentgh and feedback channel, which are neglected in this paper. For the rest of the paper, we define the achievable rate vector at coherence block $s$ as $\br(s)=[R_{1}(s), R_2(s), ..., R_{|\boldsymbol{\mathcal{P}}|}(s)]$.

	\item \textit{Adding a data point to the dataset (lines \ref{DSx}-\ref{DS}):}
	The new data point of the sampled channel vector and the corresponding rate vector $\left\langle \mathbf{\overline{h}}_{s}, \br_{s} \right\rangle$ is added to the deep learning dataset $\cD$

	\item \textit{Data transmission stage  (line \ref{DT_2}):} After the beam training, the optimal reflection beamforming vector, $\boldsymbol{\psi}_{n^\star}$, with the highest achievable rate $n^\star=\argmax_n R_n(s)$ is used to reflect the transmitted data from the transmitter for the rest of the coherence block.

	\item \textit{Deep learning model training (line \ref{DLTrain1}):} After acquiring the data entries for all $S$ coherence blocks, the deep learning model is trained using the entire dataset $\cD$. This model learns how to map an input (the sampled channel vector $\widehat{\overline{\bh}}$) to an output (predicted achievable rate with every candidate interaction vector $\widehat{\br} = \left[ \widehat{R}_{1}, \widehat{R}_{2}, \dotsc, \widehat{R}_{|\boldsymbol{\mathcal{P}}|} \right]$), as shown in \figref{fig:NN_model}. It is worth mentioning here that while we assume that the system will switch one time to PHASE II after the deep learning model is trained, the system will need to retrain and refine the model frequently to account for the changes in the environment. 
		
\end{enumerate}
\textbf{PHASE II: Prediction phase}
Following the deep learning model training in PHASE I, the LIS leverages the trained DL model to predict the reflection beamforming vector directly from the estimated \textit{sampled} channel vector, $\widehat{\overline{\mathbf{h}}}$. As shown in Algorithm \ref{alg1}, Phase II performs the following steps for every channel coherence block.
\begin{enumerate}
	\item \textit{Sampled channel estimation (line \ref{Alg_H_bar_estII}):} This step is the same as the first step in PHASE I.
	\item \textit{Achievable rate prediction (line \ref{Pred}):} In this step, the estimated \textit{sampled} channel vector, $\widehat{\overline{\mathbf{h}}}$, is fed into the deep learning model to predict the achievable rate vector, $\widehat{\br}$.
	\item \textit{Data transmission (line  \ref{DT_2_II}):} In this step, the predicted deep learning reflection beamforming vector, $\boldsymbol{\psi}_{n^\rm{DL}}$, that corresponds to the highest predicted achievable rate, is used for reflecting the transmitted data (signal). Note that instead of selecting only the interaction vector with the highest predicted achievable rate, the LIS can generally select the $k_\rm{B}$ beams corresponding to the $k_\rm{B}$ highest predicted achievable rates. It can then refine this set of beams online with the receiver to select the one with the highest achievable rate. We evaluate the performance gain if more than one reflection beam are selected in \sref{subsec:Tuning}.
	
\end{enumerate}

\subsection{Deep Learning Model} \label{subsec:DL Model}
Recent years have proven deep learning to be one of the most successful machine learning paradigms \cite{DLAppMeth}. With this motivation, a deep neural network is chosen in this work to be the model with which the desired LIS interaction function is learned. In the following, the elements of this model are described.

\textbf{Input Representation:} A single input to the neural network model is defined as a stack of environment descriptors (sampled channel vectors $\widehat{\overline{\bh}}_k$) obtained from a pair of transmitter and receiver at $K$ different sub-carrier frequencies. This sets the dimensionality of a single input vector to $K \overline{M}$. 
A common practice in machine learning is the normalization of the input data. This guarantees a stable and meaningful learning process \cite{EffBP}. The normalization method of choice here is a simple per-dataset scaling; all samples are multiplied by a factor that is the inverse of the maximum absolute value over the whole input data
\begin{equation}\label{normInput}
  \widehat{\overline{\bh}}_\rm{norm} = \frac{1}{\Delta}\widehat{\overline{\bh}},
\end{equation}
where $\Delta$ is given by
\begin{equation} \label{MaxAbs}
\Delta = \max_{\substack{ {b = 1, \dotsc, K \overline{M}}} } \left| \widehat{\overline{h}}_{b} \right|,
\end{equation} 
and $\widehat{\overline{h}}_{b}$ is the $b^{th}$ complex entry of $\widehat{\overline{\bh}}$. Besides helping the learning process, this normalization choice preserves distance information encoded in the environment descriptors. This way the model learns to become more aware of the surroundings, which is the bedrock for proposing a machine-learning-powered LIS.

The last pre-processing step of input data is to convert them into real-valued vectors without losing the imaginary-part information. This is done by splitting each complex entry into real and imaginary values, doubling the dimensionality of each input vector. The main reason behind this step is the modern implementations of DL models, which mainly use real-valued computations. 

\textbf{Target Representation:} The learning approach used in this work is supervised learning. This means the model is trained with input data that are accompanied by their so-called \textit{target responses} \cite{Goodfellow2016}. They are basically the desired responses the model is expected to approximate when it encounters inputs like those in the input training data. Since the target of the training process is to learn a function mapping descriptors to reflection vectors, the model is designed to output a set of predictions on the achievable rates of every possible reflection beamforming vector in the codebook $\left|\boldsymbol{\mathcal{P}}\right|$. Hence, the training targets are real-valued vectors, $\br(s), s=1, ..., S$, with the desired rate for each possible reflection vector.

For the same training-efficiency reason expressed for the input representation, the labels are usually normalized. The normalization used in this work is pre-sample scaling where every vector of rates $\br(s)$ is normalized using its maximum rate value $\max_n \left[\br(s)\right]_n$. The choice of normalizing each vector independently guards the model against being biased towards some strong responses. In terms of our LIS application, it gives the receivers equal importance regardless of how close or far they are from the LIS. 

\begin{figure}[t] 
	\includegraphics[width = \linewidth]{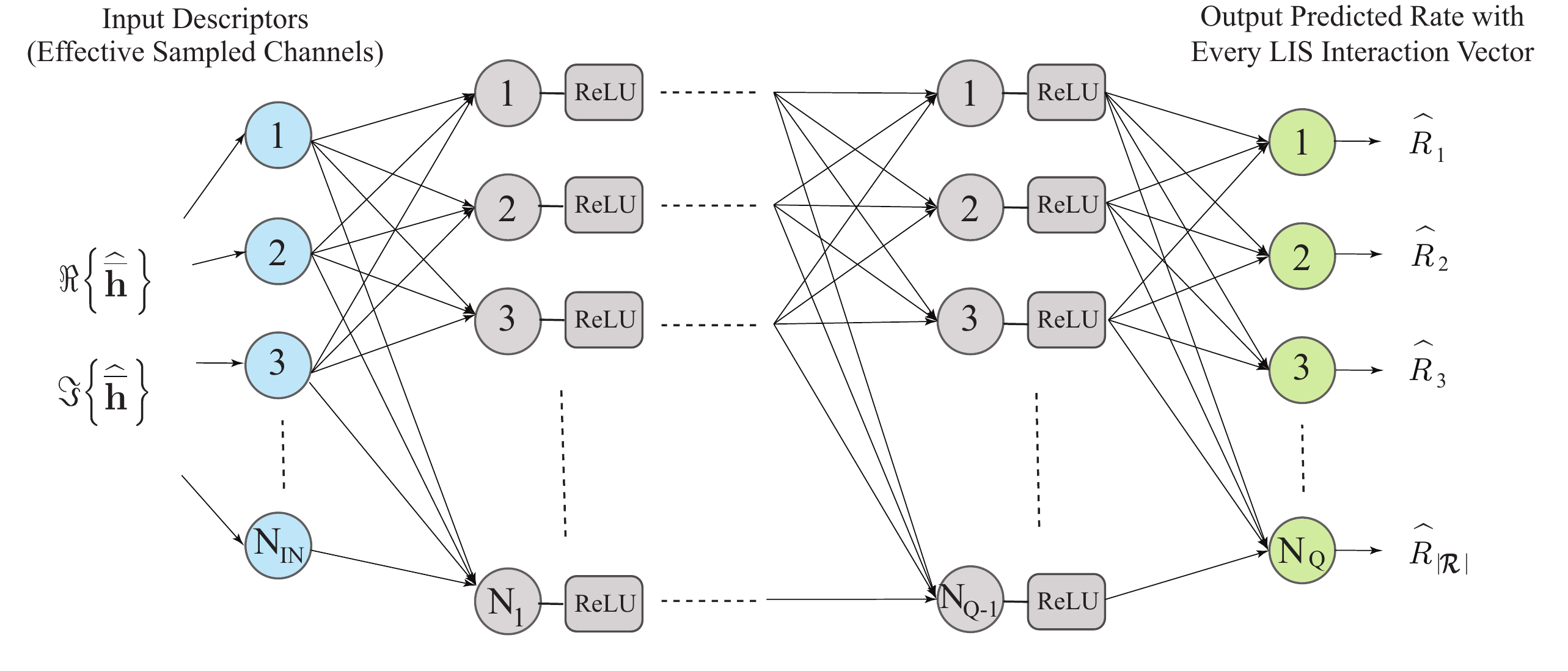}
	\caption{The adopted neural network architecture consists of $Q$ fully-connected layers. Each layer is followed by a non-linear ReLU activation layer. The deep learning model learns  how to map the observed sampled channel vectors to the predicted achievable rate using every LIS interaction vector.}
	\label{fig:NN_model}
\end{figure}

\textbf{Neural Network Architecture:} The DL model is designed as a Multi-Layer Perceptron (MLP) network, sometimes referred to as feedforward Fully Connected network. It is well-established that MLP networks are universal function approximators \cite{UniApprox}. This motivates adopting an MLP network to capture the relation between the environment descriptors and the LIS interaction (reflection beamforming) vectors. 

The proposed MLP model consists of $Q$ layers, as illustrated in \figref{fig:NN_model}. The first $Q-1$ of them alternate between fully-connected and non-linearity layers and the last one (output layer) is a fully-connected layer. The $q^{th}$ layer in the network has a stack of $N_q$ neurons, each of which sees all the outputs of the previous layer. For the non-linearity layers, they all employ Rectified Linear Units (ReLUs) \cite{Goodfellow2016}. Each unit operates on a single input value outputting another single value. Hence, the number of units per layer equals the number of outputs of the previous fully-connected layer.

\textbf{Training Loss Function:} The model training process aims at minimizing a loss function that measures the quality of the model predictions. Given the objective of predicting the best reflection beam vector, $\boldsymbol{\psi}_{n^\mathrm{DL}}$, having the highest achievable rate estimate, $\max_n \widehat{R}_n$, the model is trained using a regression loss function. At every coherence block, the neural network is trained to make its output, $\widehat{\br}$, as close as possible to the desired output, the normalized achievable rates, $\overline{\br}$. Specifically, the training is guided through minimizing the loss function, $L\left(\boldsymbol{\theta}\right)$, expressed as
\begin{equation} \label{eq:24}
L\left(\boldsymbol{\theta}\right)= \mathsf{MSE} \left( \overline{\br}, \widehat{\br}\right),
\end{equation}
where $\boldsymbol{\theta}$ represents the set of all the neural network parameters and $\mathsf{MSE} \left( \overline{\br}, \widehat{\br} \right)$ indicates the mean-squared-error between $\overline{\br}$ and $\widehat{\br}$.

\section{Simulation Results}  \label{sec:Results}

In this section, we evaluate the performance of both the deep learning (DL) and the compressive sensing (CS) based reflection beamforming approaches.  The flow of this section is as follows. First, we describe the adopted experimental setup and datasets.  Then, we compare the performance of the deep learning and compressive sensing solutions at both mmWave and sub-6 GHz bands. After that, we investigate the impact of different system and machine learning parameters on the performance of the deep learning solution.

\begin{figure}[t] \centerline{\includegraphics[scale=.3]{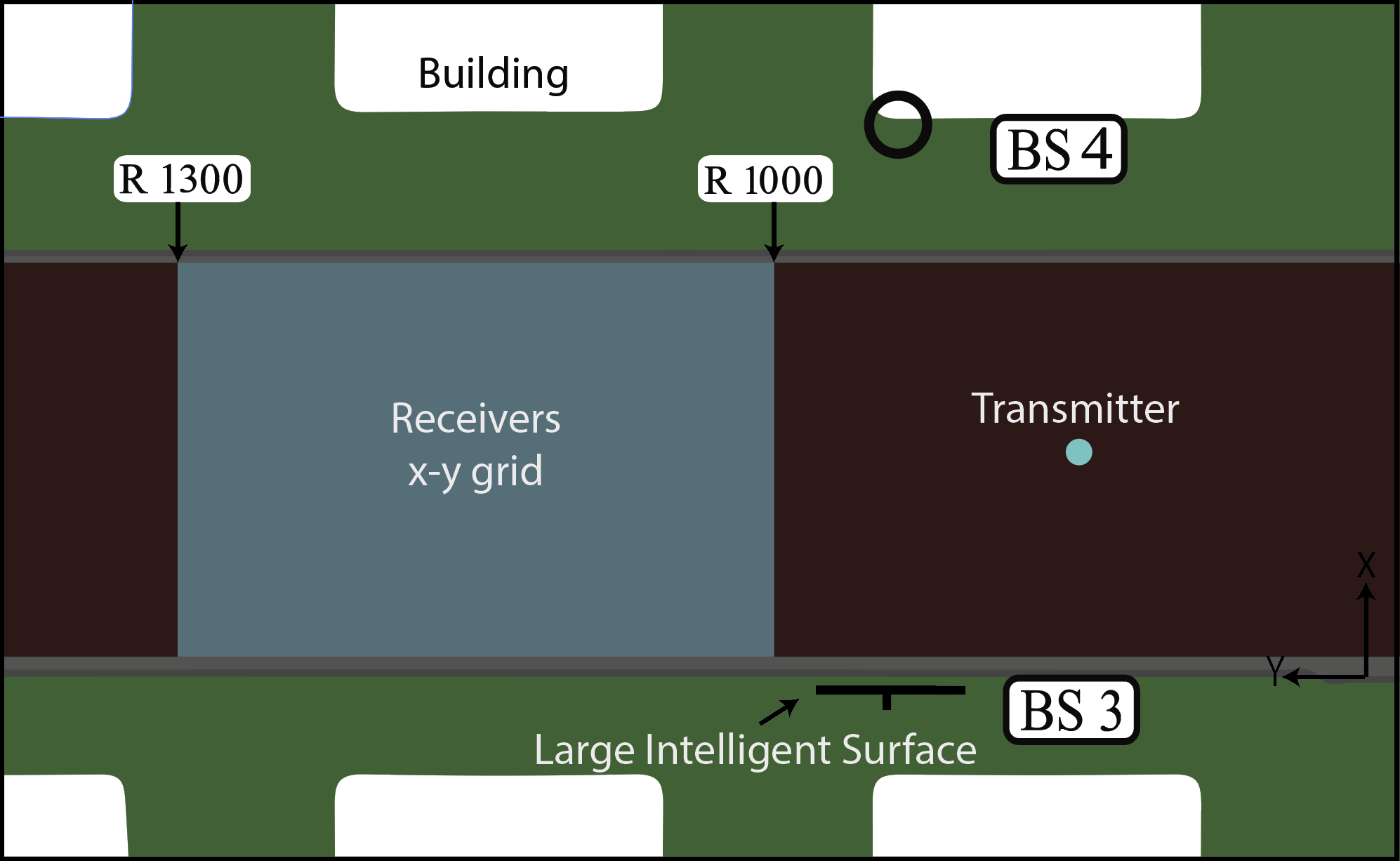}}
	\caption{This figure illustrates the adopted ray-tracing scenario where an LIS is reflecting the signal received from one fixed transmitter to a receiver. The receiver is selected from an x-y grid of candidate locations. This ray-tracing scenario is generated using Remcom Wireless InSite \cite{Remcom}, and is publicly available on the DeepMIMO dataset \cite{DeepMIMO2019}. }
	\label{fig:Setup}
\end{figure}

\subsection{Simulation Setup} \label{subsec:Sim_Setup}

Given the geometric channel model adopted in \sref{sec:SysCh Model} and the nature of the reflection beamforming optimization problem, with its strong dependence on the environmental geometry, it is critical to evaluate the performance of the proposed solutions based on realistic channels. This motivates using channels generated by ray-tracing to capture the  dependence on the key environmental factors such as the environment geometry and materials, the LIS and transmitter/receiver locations,  the operating frequency, etc. To do that, we adopted the DeepMIMO dataset, described in detail in \cite{DeepMIMO2019}, to generate the channels based on the outdoor ray-tracing scenario `O1' \cite{Remcom}, as will be discussed shortly. The DeepMIMO  is a parameterized dataset published for deep learning applications in mmWave and massive MIMO systems. The machine learning simulations were executed using the Deep Learning Toolbox of MATLAB R2019a. Next, we explain in detail the key components of the simulation setup.

\begin{table}[h]
	\caption{The adopted DeepMIMO dataset parameters}
	\begin{center}
		\begin{tabular}{ | c | c | }
			\hline
			\textbf{DeepMIMO Dataset Parameter} & \textbf{Value} \\ \hline \hline
			Active BSs &    $3$  \\ \hline
			Active users emulating the receivers  &  From row R1000 to row R1300  \\ \hline
			Active user emulating the transmitter  &  row R850 column 90  \\ \hline
			Number of BS Antennas & $\left(M_x, M_y, M_z\right) = \left(1,16,16 \right); \left(1,32,32 \right);\left(1,64,64\right)$ \\ \hline
			Antenna spacing & $0.5$ \\ \hline 
			System bandwidth & $100$ MHz \\ \hline 
			Number of OFDM subcarriers & $512$ \\ \hline
			OFDM sampling factor & $1$ \\ \hline
			OFDM limit & $64$ \\ \hline 
			Number of paths & $1, 2, 5, 10$ \\ \hline
		\end{tabular}
	\end{center}
	\label{table:DeepMIMOpset}
\end{table}

\textbf{System model:}  We adopt the system model described in \sref{sec:Sys_Model} with one large intelligent surface that reflects the signal received from a transmitter to a receiver. The transmitter is assumed to be fixed while the receiver can take any random position in a specified x-y grid as illustrated in \figref{fig:Setup}. We implemented this setup using the outdoor ray-tracing scenario 'O1' of the DeepMIMO dataset that is publicly available at \cite{DeepMIMO2019}.  As shown in \figref{fig:Setup}, we select BS 3 in the 'O1' scenario to be the LIS and the user in row R850 and column 90 to be the fixed transmitter. The uniform x-y grid of candidate receiver locations include $54300$ points from row R1000 to R1300 in the 'O1' scenario where every row consists of 181 points. Unless otherwise mentioned, the adopted LIS employs a UPA with 64x64 ($M=4096$) antennas at the mmWave 28GHz setup and a UPA with 16x16 ($M=256$) antennas at the 3.5GHz setup. The active channel sensors described in \sref{sec:LIS_arch} are randomly selected from the $M$ UPA antennas. The transmitter and receiver are assumed to have a single antenna each. The antenna elements have a gain of 3dBi and the transmit power is 35dBm. The rest of the adopted DeepMIMO dataset parameters are summarized in Table \ref{table:DeepMIMOpset}.

\textbf{Channel generation:}
The channels between the LIS and the transmitter/receiver, $\bh_{\rm{T},k}, \bh_{\rm{R},k}$, for all the candidate receiver locations in the x-y grid, are constructed using the DeepMIMO dataset generation code \cite{DeepMIMO2019} with the parameters in Table \ref{table:DeepMIMOpset}.  With these channels, and given the randomly selected active elements in the proposed LIS architecture, we construct the sampled channel vectors $\overline{\bh}_{\rm{T},k}, \overline{\bh}_{\rm{R},k}$. The noisy sampled channel vectors $\widehat{\overline{\bh}}_{\rm{T},k}, \widehat{\overline{\bh}}_{\rm{R},k}$ are then generated by adding noise vectors to $\overline{\bh}_{\rm{T},k}, \overline{\bh}_{\rm{R},k}$ according to \eqref{eq:noisy_ch}, with the noise power calculated based on the bandwidth and other parameters in Table \ref{table:DeepMIMOpset}, and with receiver noise figure of $5$dB. These noisy sampled channels are then used to design the LIS interaction (reflection beamforming) vectors following the proposed compressive sensing and deep learning approaches.

\textbf{LIS interaction (reflection beamforming) codebook:} We adopt a DFT codebook for the candidate LIS interaction vectors. More specifically, considering the UPA structure, we define the LIS interaction codebook as $\rm{DFT}_{M_\rm{H}} \kron \rm{DFT}_{M_\rm{V}}$. The codebook $\rm{DFT}_{M_\rm{H}} \in \mathbb{C}^{{M_\rm{H}}  \times {M_\rm{H}}}$ is a DFT codebook for the azimuth (horizontal) dimension where the $m_\rm{H}$th column, $m_\rm{H}=1,2, ...,M_\rm{H}$, is defined as $[1, e^{-j \frac{2 \pi}{M_\rm{H}} m_\rm{H}},  ..., e^{-j (M_\rm{H}-1) \frac{2 \pi}{M_\rm{H}}  m_\rm{H} } ]^T$. The codebook $\rm{DFT}_{M_\rm{V}}$ is similarly defined.

\textbf{Deep learning parameters:} We adopt the deep learning model described in \sref{subsec:DL Model}. To reduce the neural network complexity, however, we input the normalized sampled channels only at the first $K_\mathrm{DL}=64$ subcarriers, where $K_\mathrm{DL} \leq K$, which sets the length of the DL input vector to be $2\overline{M} K_\mathrm{DL}$. The length of the DL output vector is $M=|\boldsymbol{\mathcal{P}}|$, as described in \sref{subsec:DL Model}. The neural network architecture consists of four fully connected layers. The number of hidden nodes of the four layers are $M, 4 M, 4 M, M$, respectively, where $M$ is the number of LIS antennas. Given the size of the x-y grid of the candidate receiver locations in \figref{fig:Setup}, the deep learning dataset has $54300$ data points. We split this dataset into two sets, namely a training set and a testing set with $85\%$ and $15\%$ of the points, respectively. Unless otherwise mentioned, we consider a batch size of $500$ samples and a $50 \%$ dropout rate. The dropout layer is added after every ReLU  layer.

\begin{figure}[t] \centerline{\includegraphics[width=0.75\columnwidth,height=260pt]{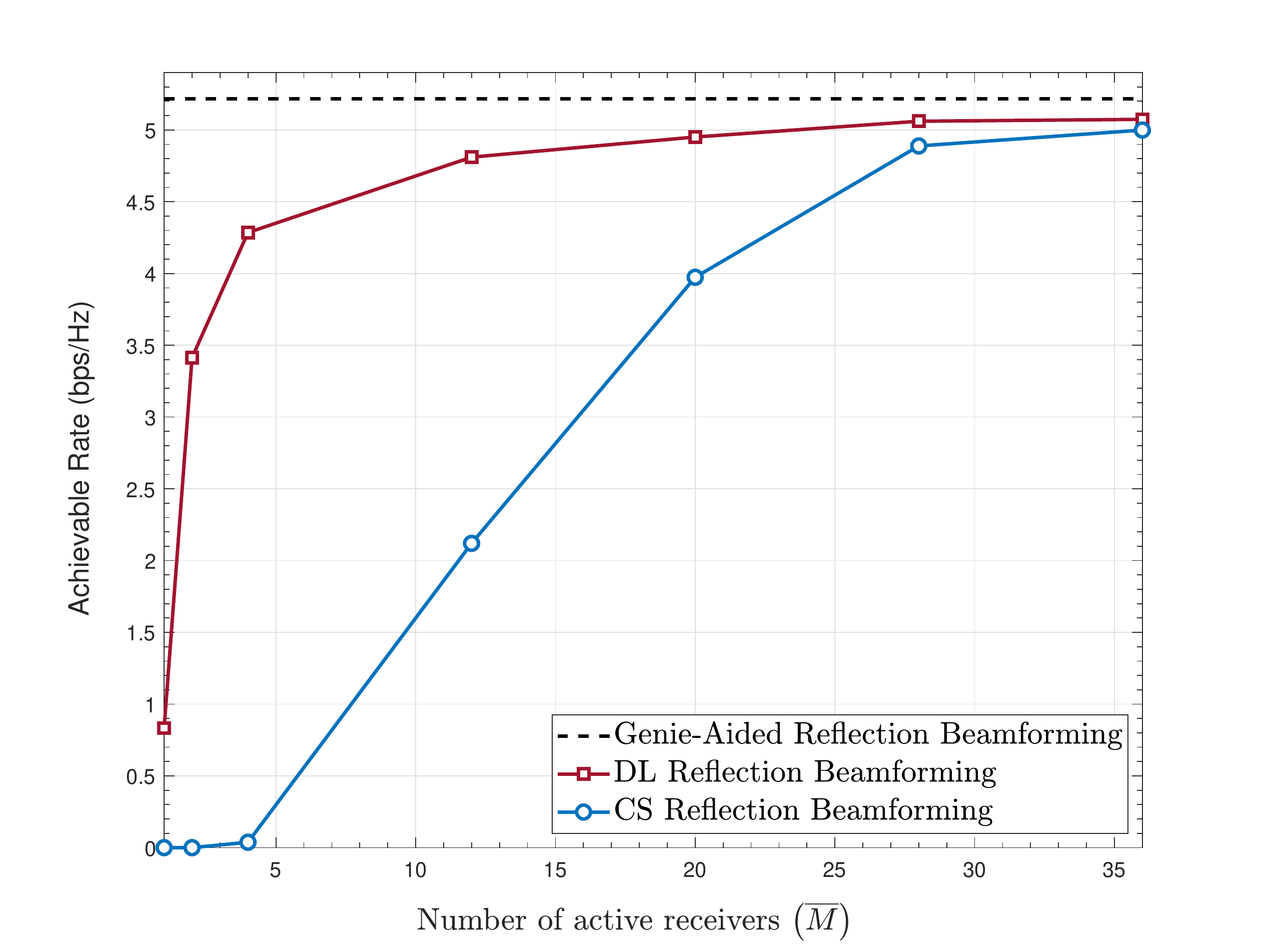}}
	\caption{The achievable rate of both proposed CS and DL based reflection beamforming solutions are compared to the upper bound $R^\star$ and the CS based beamforming approach, for different numbers of active receivers, $\overline{M}$. The figure is generated at $f_c = 28$ GHz, $M=64 \times 64$ antennas, and $L=10$ paths.} 
	\label{fig:Sim8b}
\end{figure}

\textbf{Compressive sensing parameters:} We consider the developed compressive sensing solution in \sref{sec:BL_Solution} to recover the full LIS-transmitter/receiver channels and design the LIS reflection beamforming vectors. For approximating the solution of \eqref{eq:sparse_h}, we use OMP with a grid of size $N_\rm{D}^\rm{Az} N_\rm{D}^\rm{El}$ points, where $N_\rm{D}^\rm{Az}=2 M_\rm{H}, N_\rm{D}^\rm{El}=2 M_\rm{V}$.

Next, we evaluate the performance of the developed compressive sensing and deep learning solutions for designing the LIS interaction vectors. 

\begin{figure}[t] \centerline{\includegraphics[width=0.75\columnwidth,height=260pt]{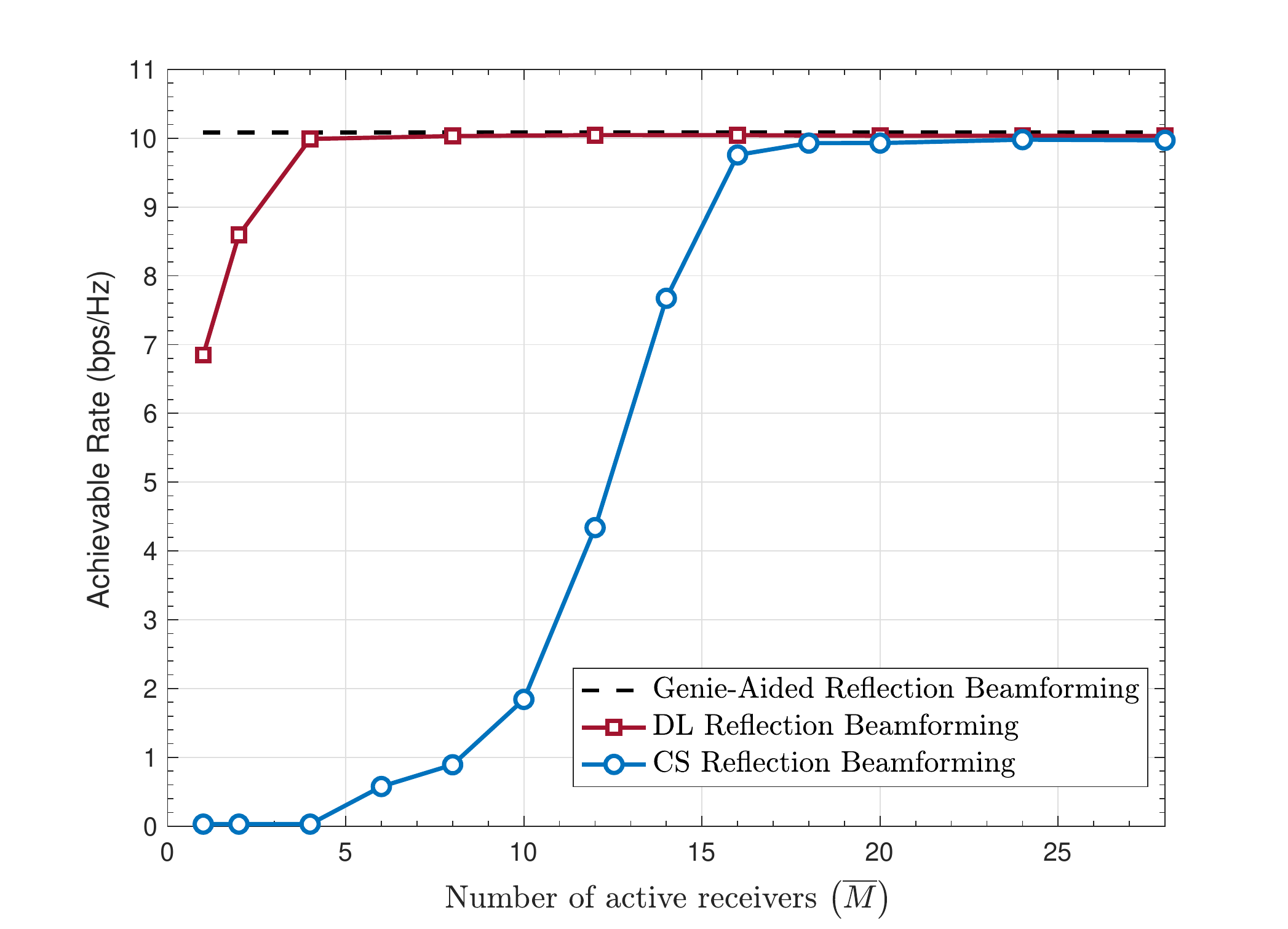}}
	\caption{The achievable rate of the proposed DL based reflection beamforming approach compared to the upper bound $R^\star$ and the CS based beamforming approach, for different numbers of active receivers, $\overline{M}$. The figure is generated at $f_c = 3.5$ GHz, $M=16 \times 16$ antennas,  and $L=15$ paths.} 
	\label{fig:Sim8a}
\end{figure}

\subsection{Achievable Rates with Compressive Sensing and Deep Learning Based LIS Systems} \label{subsec:Compare_CS_DL}

In this subsection, we evaluate the achievable rates of the proposed compressive sensing (CS) and deep learning (DL) based reflection beamforming solutions for LIS systems. These rates are compared to the genie-aided upper bound, $R^\star$, in \eqref{eq:optR} which assumes perfect knowledge of the full channel vectors $\bh_{\rm{T},k}, \bh_{\rm{R},k}$. In \figref{fig:Sim8b}, we consider the simulation setup in \sref{subsec:Sim_Setup} at the mmWave 28GHz band with LIS employing a UPA of $64 \times 64$ antennas. The channels are constructed to include the strongest  $L=10$ channel paths. Further, the LIS employs the proposed architecture in \figref{fig:Sys_Model_Sparse} with only $\overline{M}$ active sensors. \figref{fig:Sim8b} shows that the proposed deep learning solution approaches the optimal upper bound with a very small number of active antennas. For example, with only $\overline{M}=4$  active antennas (out of $M=4096$ total antennas), the  deep learning solution achieves almost $85\%$ of the optimal achievable rate. This figure also illustrates the performance gain of the deep learning solution compared to the compressive sensing approach, especially when the number of active antennas is very small.  Note that the two solutions  approach the upper bound with $28-36$ active antennas, which represent less than $1\%$ of the total number of antennas ($M=4096$) in the LIS. This illustrates the high energy efficiency of the proposed LIS architecture and reflection beamforming solutions. 

\begin{figure}[t] \centerline{\includegraphics[width=0.75\columnwidth,height=260pt]{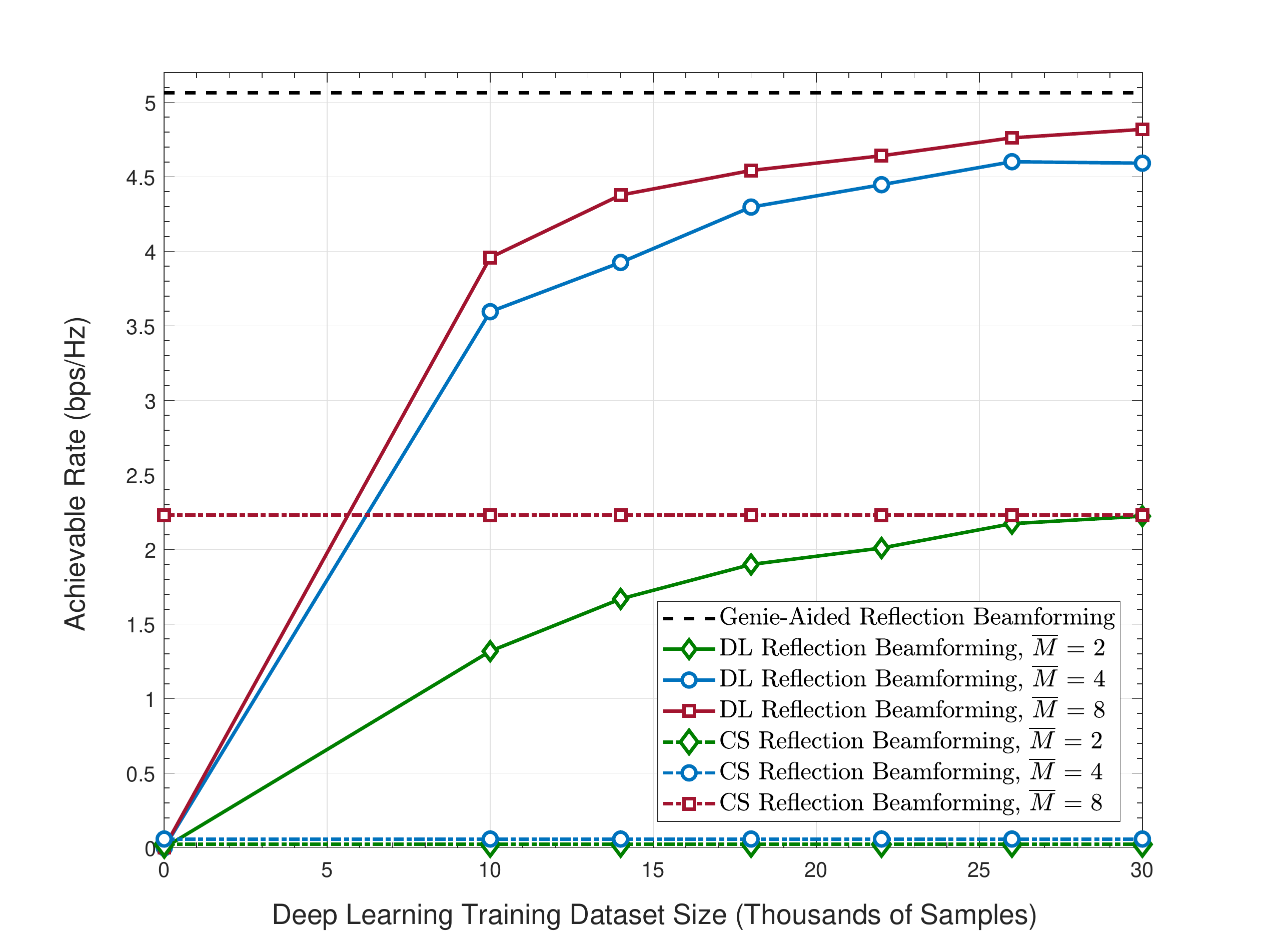}}
	\caption{The achievable rate of the proposed DL based reflection beamforming approach is compared to the upper bound $R^\star$ and the CS beamforming approach, for different numbers of active receivers, $\overline{M}$. The adopted setup considers an LIS with $64 \times 64$ UPA. This figure highlights the promising gain of the proposed deep learning solution that approaches the upper bound using only $8$ active elements (less than $1\%$ of the total number of antennas). This performance requires collecting a dataset of around 20-25 thousand data points (user locations).} 
	\label{fig:Sim1}
\end{figure}

To evaluate the performance at sub-6 GHz systems, we plot the achievable rates of the proposed deep learning and compressive sensing solutions compare to the optimal rate $R^\star$ in \figref{fig:Sim8a}. This figure adopts the simulation setup in \sref{subsec:Sim_Setup} at a $3.5$GHz band. The LIS is assumed to employ a UPA with $16 \times 16$ antennas and each channel incorporates the strongest $L=15$ paths. \figref{fig:Sim8a} shows that the proposed deep learning and compressive sensing solutions are also promising for sub-6 GHz LIS systems. This is captured by the convergence to the upper bound with only $4$ active elements in the deep learning case and around $18$ elements in the compressive sensing case. This figure also illustrates the larger gain of the deep learning solution compared to the compressive sensing approach in the sub-6 GHz systems, where the channels are less sparse than mmWave systems. This gain, however, has the cost of collecting a dataset to train the  deep learning model, which is not required in the compressive sensing approach.

\begin{figure}[t] \centerline{\includegraphics[width=0.75\columnwidth,height=260pt]{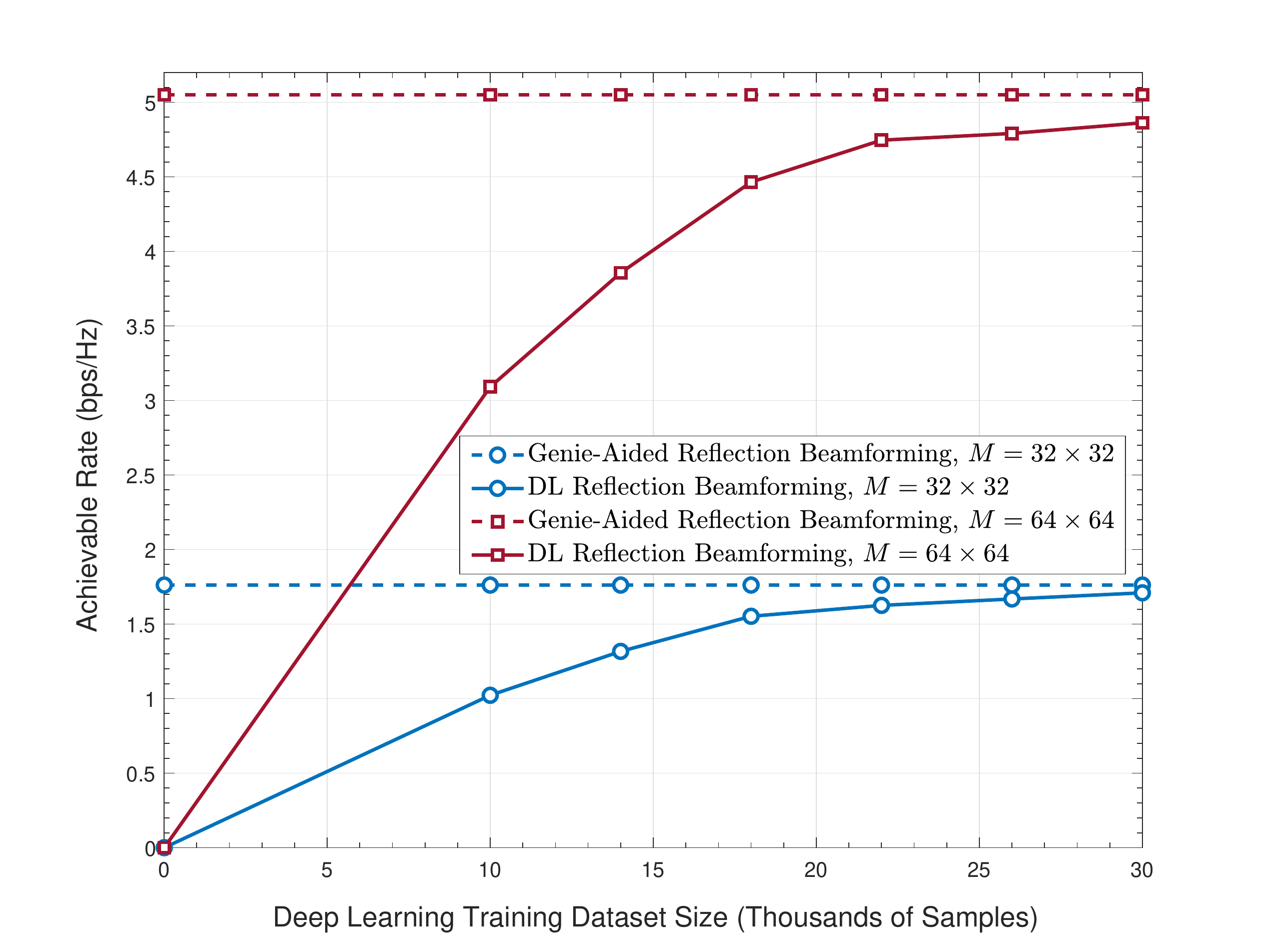}}
	\caption{The achievable rate of the proposed DL based reflection beamforming approach is compared to the upper bound $R^\star$ for different sizes of intelligent surfaces, namely with LIS of $32 \times 32$ and $64 \times 64$ UPAs. The number of active elements (channel sensors) equals $\overline{M}=8$. } 
	\label{fig:Sim2}
\end{figure}

\subsection{How much training is needed for the deep learning model?} \label{subsec:Efficiency}

The data samples in the deep learning dataset are captured when the receiver is randomly sampling the x-y grid. In \figref{fig:Sim1}, we study the performance of the developed deep learning approach for designing the LIS interaction vectors for different dataset sizes. This illustrates the improvement in the machine learning prediction quality as it sees more data samples. For \figref{fig:Sim1}, we adopt the simulation setup in \sref{subsec:Sim_Setup} with an LIS of $64 \times 64$ UPA and a number of active channel sensors $\overline{M}=2,4$, and $8$. The setup considers a mmWave $28$GHz scenario and the channels are constructed with only the strongest path, i.e., $L=1$. \figref{fig:Sim1} shows that with only $8$ active antennas, the proposed deep learning solution can achieve almost $90\%$ of the optimal rate in \eqref{eq:optR} when the model is trained on $10$ thousand data points (out of the 53400 points) in the x-y grid. Further, this figure highlights the performance gain of the deep learning approach compared to the compressive sensing solution. This gain increases with larger dataset sizes as the compressive sensing solution does not leverage the prior channel estimation/LIS interaction observations and its performance does not depend on the size of the dataset.

\subsection{Impact of Important System and Channel Parameters} \label{subsec:Impact}

\begin{figure}[t] \centerline{\includegraphics[width=0.75\columnwidth,height=260pt]{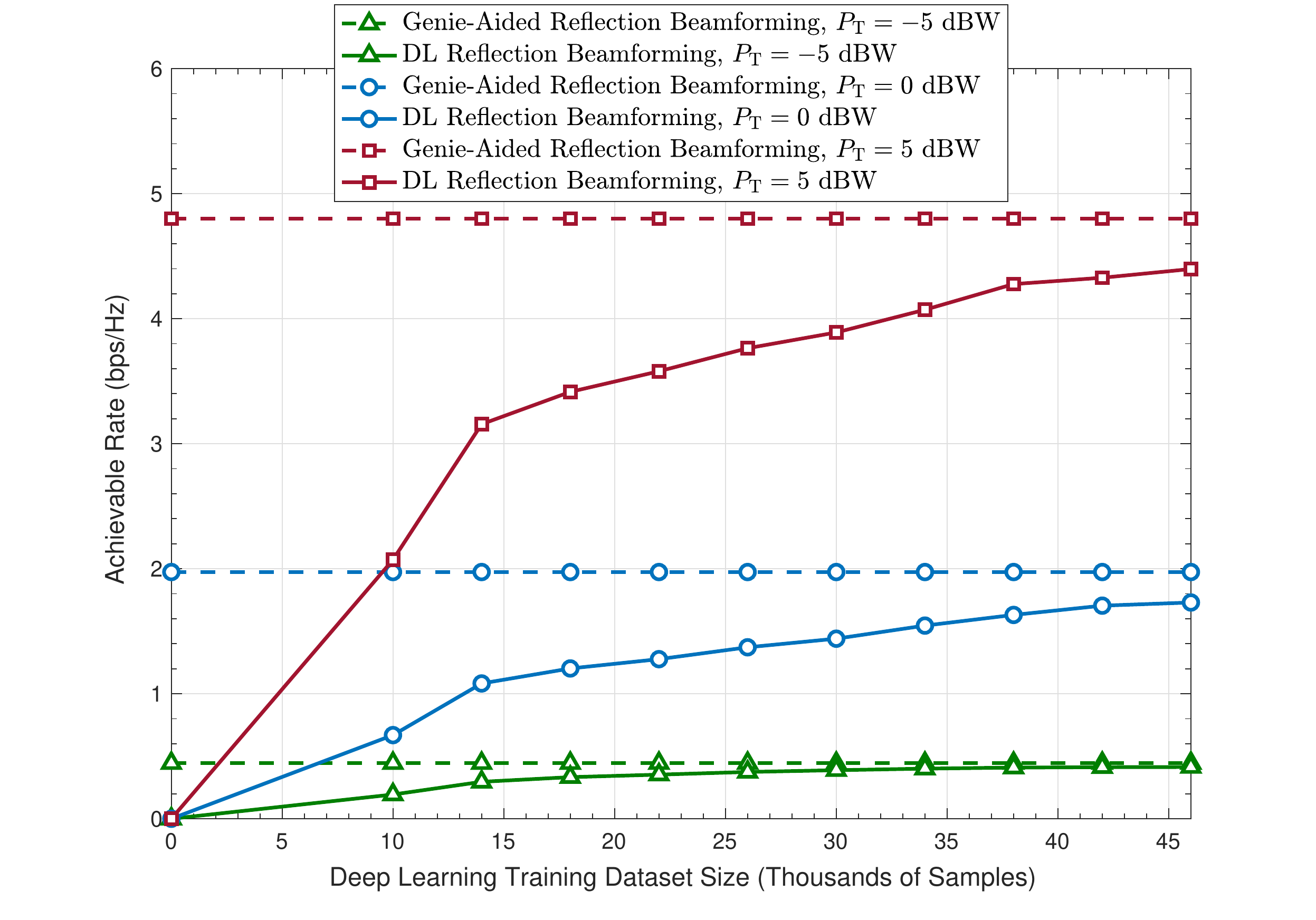}}
	\caption{The achievable rate of the proposed deep learning based reflection beamforming approach is compared to the upper bound $R^\star$, for different values of user transmit power, $P_\rm{T}$. The figure is generated for an LIS with $M=64 \times 64$ UPA and $\overline{M}=8$ active elements. This figure shows that the proposed DL solution is capable of learning and approaching the optimal achievable rate even with a relatively small transmit power.} 
	\label{fig:Sim4}
\end{figure}

In this subsection, we evaluate the impact of the key system and channel parameters on the performance of the developed deep learning solution.  

\textbf{Number of LIS antennas:}
\figref{fig:Sim2} examines the achievable rate performance of the developed solutions for designing the LIS interaction vectors when the LIS employs either a $32 \times 32$ or a $64 \times 64$ UPA. This figure adopts the same mmWave scenario considered in \figref{fig:Sim1}. As illustrated, with only $\overline{M}=8$ active receivers, the proposed deep learning solution approaches the optimal rate in \eqref{eq:optR} that assumes perfect channel knowledge for different LIS sizes. This shows the potential of the proposed LIS architecture and deep learning solution in enabling large intelligent surfaces with large numbers of antennas. \textbf{Note that the proposed solution does not requires any beam training overhead (as it relies on the deep learning prediction of the best beam) and needs only $8$ active receivers to realize this near-optimal performance in \figref{fig:Sim2}}.

\textbf{Transmit power:}
In \figref{fig:Sim4}, we study the impact of the transmit power (and receive SNR) on the achievable rate performance of the developed deep learning solution. This is important in order to evaluate the robustness of the learning and prediction quality, as we input the noisy sampled channel vectors to the deep learning model. In \figref{fig:Sim4}, we plot the achievable rates of the proposed deep learning solution as well as the upper bound in \eqref{eq:optR} for three values of the transmit power, $P_\rm{T}=-5, 0, 5$ dBW. These transmit powers map to \textit{receive} SNR values of $-3.8, 6.2, 16.2$ dB, respectively, including the LIS beamforming gain of the $4096$ antennas. The rest of the setup parameters are the same as those adopted in \figref{fig:Sim1}. \figref{fig:Sim4} illustrates that the proposed deep learning solution can perform well even with relatively small transmit powers and low SNR regimes. 

\begin{figure}[t] \centerline{\includegraphics[width=0.75\columnwidth,height=260pt]{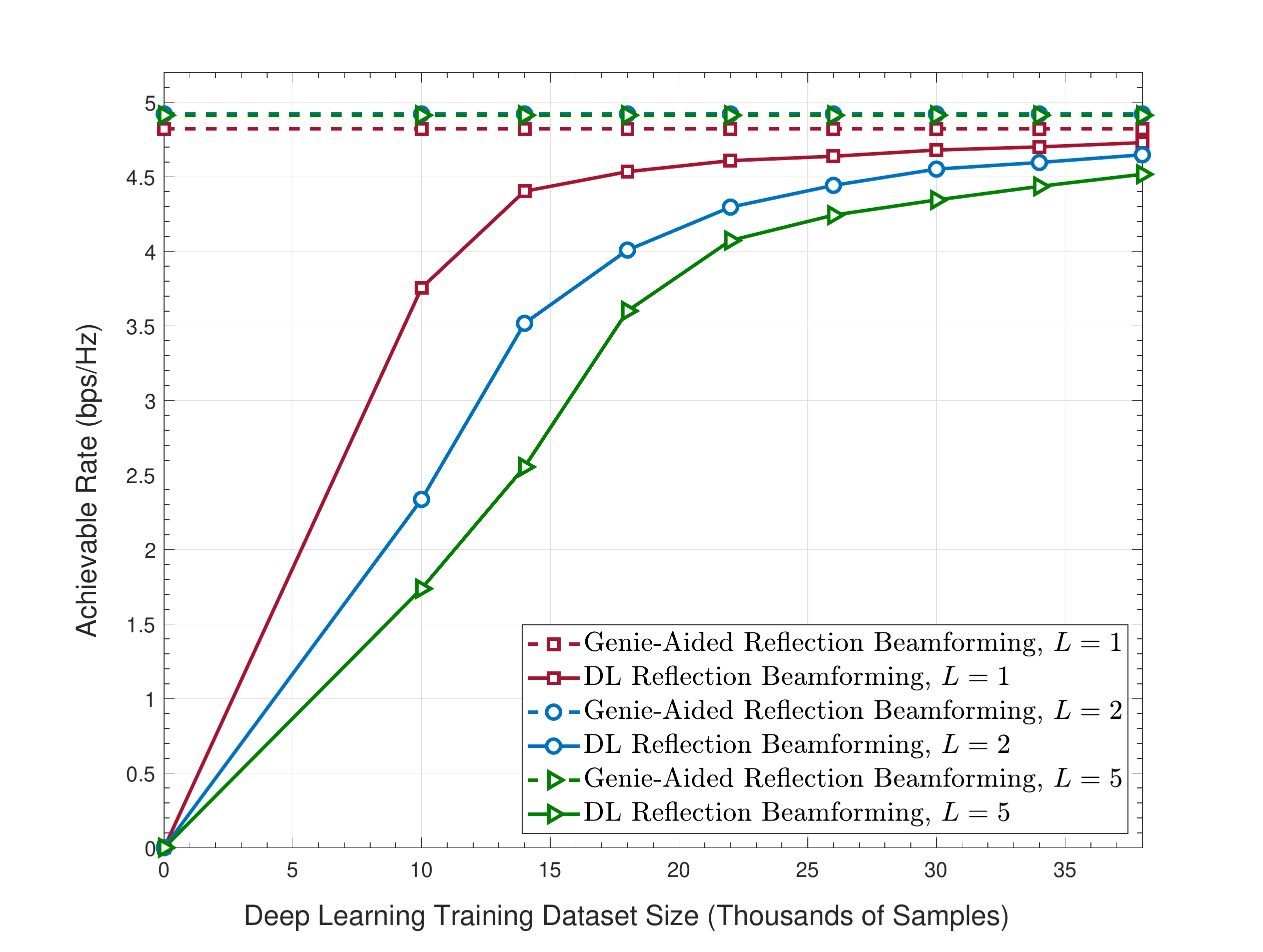}}
	\caption{The achievable rate of the proposed DL based reflection beamforming approach is compared to the upper bound $R^\star$, for different numbers of channel paths, $L$. The figure is generated for an LIS with $64 \times 64$ UPA and $\overline{M}=4$ active elements. As the number of channel paths increases, the achievable rate achieved by the proposed DL solution converges slower to the upper bound. Hence, using more training data can help learn multi-path signatures. } 
	\label{fig:Sim6}
\end{figure}

\textbf{Number of channel paths:} 
In \figref{fig:Sim6}, we investigate the impact of the number of channel paths on the performance of the developed deep learning solution. In other words, we examine the robustness of the proposed deep learning model with multi-path channels. For this figure, we adopt the same simulation setup of \figref{fig:Sim1} with an LIS employing $64 \times 64$ UPA. The channels are constructed considering the strongest $L=1, 2$, or $5$ channel paths. As illustrated in \figref{fig:Sim6}, with the increase in the number of channel paths, the achievable rate by the proposed deep learning solution converges slower to the upper bound. This shows that the proposed deep learning model can learn from multi-path channels if a large enough dataset is available.

\begin{figure}[t] \centerline{\includegraphics[width=0.75\columnwidth,height=260pt]{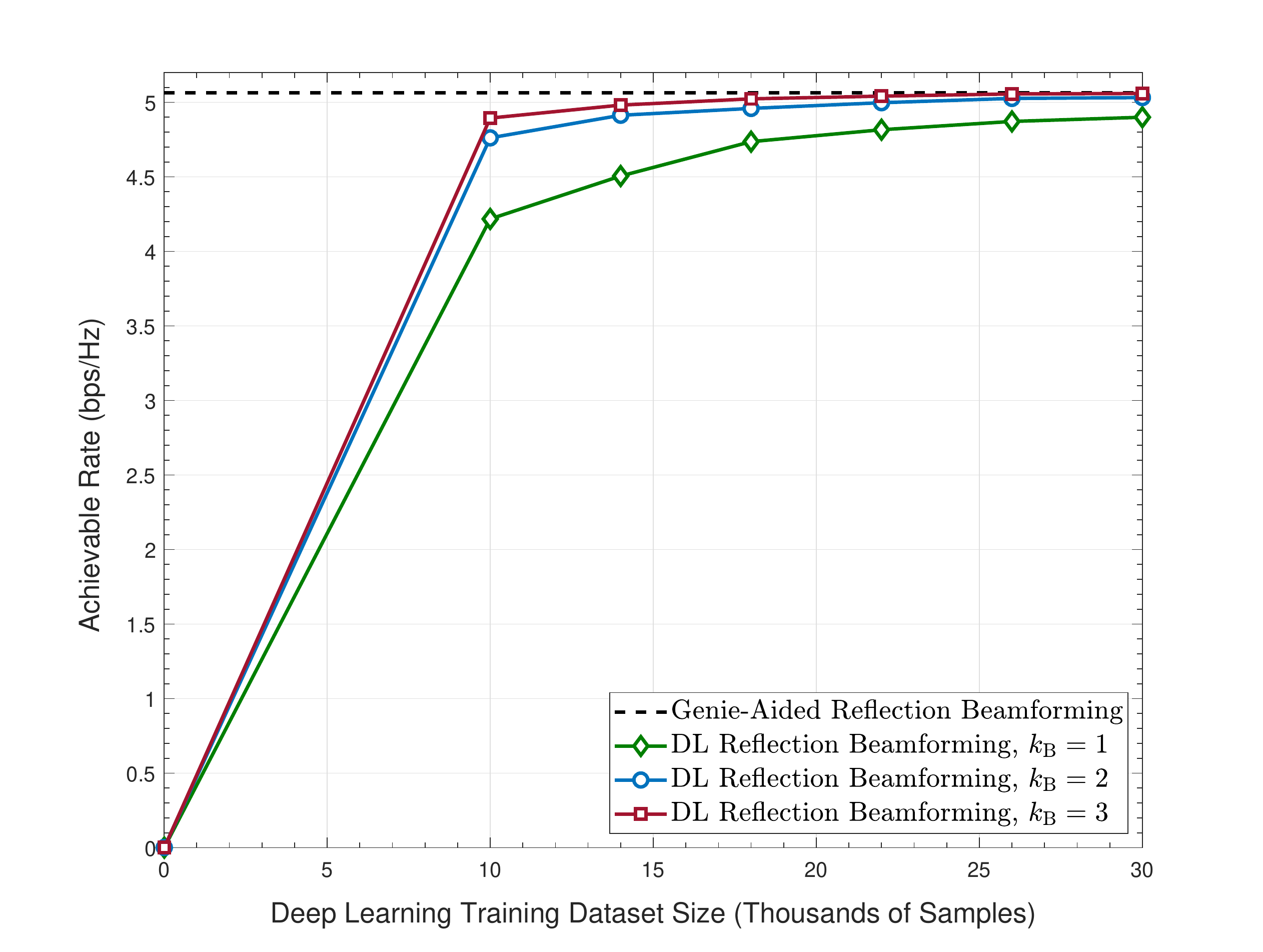}}
	\caption{The achievable rate of the proposed DL based reflection beamforming approach is compared to the upper bound $R^\star$. The simulation considers an LIS with $64 \times 64$ UPA and $\overline{M}=4$ active channel sensors. The figure illustrates the achievable rate gain when the beams selected by the deep learning model is further refined through beam training over $k_\rm{B}$ beams. }
	\label{fig:Sim5}
\end{figure}

\subsection{Refining the deep learning prediction } \label{subsec:Tuning}
In \figref{fig:Sim8b}-\figref{fig:Sim6}, we considered the proposed deep learning solution where the deep learning model use the sampled channel vectors to predict the best beam and this beam is directly used to reflect the transmitted data. Relying completely on the deep learning model to determine the reflection beamforming vector has the clear advantage of eliminating the beam training overhead and enabling highly-mobile applications. The achievable rates using this approach, however, may be sensitive to small changes in the environment. A candidate approach for enhancing the reliability of the system is to use the machine learning model to predict the most promising $k_\rm{B}$ beams. These beams are then refined through beam training with the receiver to select the final beam reflection vector. Note that the most promising $k_\rm{B}$ beams refer to the $k_\rm{B}$ beams with the highest predicted rates from the deep learning model. To study the performance using this approach, we plot the achievable rate of the deep learning solution in \figref{fig:Sim5}, for different values of $k_\rm{B}$. As this figure shows, refining the most promising $k_\rm{B}$ yields higher achievable rates compared to the case when the LIS relies completely on the deep learning model to predict the best beam, i.e., with $k_\rm{B}$. The gain in \figref{fig:Sim5} is expected to increase with more time-varying and dynamic environment, which is an interesting extension in the future work.

\section{Conclusion}  \label{sec:Conclusion}

In this paper, we considered LIS-assisted wireless communication systems and developed efficient solutions that design the LIS interaction (reflection) matrices with negligible training overhead. More specifically, we first introduced a novel LIS architecture where only a small number of the LIS elements are active (connected to the baseband). Then, we developed two solutions that design the LIS reflection matrices for this new architecture with almost no training overhead. The first solution leverages compressive sensing tools to construct the channels at all the antenna elements from the \textit{sampled} channels seen only at the active elements. The second approach exploits deep learning tools to learn how to predict the optimal LIS reflection matrices directly from the sampled channel knowledge, which represent what we call \textit{environment descriptors}. Extensive simulation results based on accurate ray-tracing showed that the two proposed solutions can achieve near-optimal data rates with negligible training overhead and with a small number of active elements. Compared to the compressive sensing solution, the deep learning approach requires a smaller number of active elements to approach the optimal rate thanks to leveraging its prior observations. Further, the deep learning approach does not require any knowledge of the LIS array geometry and does not assume sparse channels. To achieve these gains, however, the deep learning model needs to collect enough dataset, which is not needed in the compressive sensing solution. For the future work, it is interesting to investigate other machine learning models such as  the use of reinforcement learning that does not require an initial dataset collection phase. For the compressive sensing solution, there are several interesting extensions, including the optimization of the sparse distribution of the active sensors leveraging tools from nested and co-prime arrays.

\linespread{1.2}


\end{document}